\documentclass[aps, prd, floatfix, nofootinbib, superscriptaddress, twocolumn]{revtex4-2}

\usepackage{amsmath}
\usepackage{amssymb}
\usepackage{amsfonts}
\usepackage{color}
\usepackage{CJKutf8}
\usepackage{latexsym}
\usepackage{textcomp}

\usepackage[mathscr,scaled=1.15]{urwchancal}
\DeclareFontFamily{OT1}{pzc}{}
\DeclareFontShape{OT1}{pzc}{m}{it}%
{<-> s * [1.15] pzcmi7t}{}
\DeclareMathAlphabet{\mathpzc}{OT1}{pzc}{m}{it}

\usepackage{supertabular}
\usepackage{placeins}
\usepackage{epsfig}
\usepackage{graphicx}
\usepackage{multirow}

\definecolor{purple}{rgb}{0.5,0,0.5}
\definecolor{blue}{rgb}{0.0,0,0.9}
\definecolor{prdblue}{rgb}{0.133,0.118,0.498}
\usepackage[colorlinks=true, pdfstartview=FitV, linkcolor=prdblue, citecolor= prdblue, urlcolor=prdblue]{hyperref}

\newcounter{Afigure}
\newcounter{Atable}

\begin{document}

\begin{CJK}{UTF8}{song}

\title{$\,$\\[-6.5ex]\hspace*{\fill}{\normalsize{\sf\emph{Preprint nos}.\ NJU-INP 063/22, USTC-ICTS/PCFT-22-21}}\\[1.25ex]
Contact interaction analysis of octet baryon axialvector and pseudoscalar form factors
}

\date{2022 August 03}

\author{Peng Cheng
        $\,^{\href{https://orcid.org/0000-0002-6410-9465}{\textcolor[rgb]{0.00,1.00,0.00}{\sf ID}}}$}
\affiliation{School of Physics, Nanjing University, Nanjing, Jiangsu 210093, China}
\affiliation{Institute for Nonperturbative Physics, Nanjing University, Nanjing, Jiangsu 210093, China}
\author{Fernando E.~Serna%
        $\,^{\href{https://orcid.org/0000-0003-2032-9412}{\textcolor[rgb]{0.00,1.00,0.00}{\sf ID}}}$}
\affiliation{Departamento de Fisica, Universidad de Sucre, Carrera 28 No.\,5-267, Barrio Puerta Roja, Sincelejo, Colombia}
\affiliation{Laborat\'{o}rio de F\'{\i}sica Te\'{o}rica e Computacional, Universidade Cidade de S\~{a}o Paulo, Rua Galv\~{a}o Bueno 868, 01506-600 S\~{a}o Paulo, Brazil}
\author{Zhao-Qian Yao
        $\,^{\href{https://orcid.org/0000-0002-9621-6994}{\textcolor[rgb]{0.00,1.00,0.00}{\sf ID}}}$}
\affiliation{School of Physics, Nanjing University, Nanjing, Jiangsu 210093, China}
\affiliation{Institute for Nonperturbative Physics, Nanjing University, Nanjing, Jiangsu 210093, China}
\author{\\Chen Chen
        $\,^{\href{https://orcid.org/0000-0003-3619-0670}{\textcolor[rgb]{0.00,1.00,0.00}{\sf ID}}}$}
\email[]{chenchen1031@ustc.edu.cn}
\affiliation{Interdisciplinary Center for Theoretical Study, University of Science and Technology of China, Hefei, Anhui 230026, China}
\affiliation{Peng Huanwu Center for Fundamental Theory, Hefei, Anhui 230026, China}
%
\author{\mbox{Zhu-Fang Cui}
        $\,^{\href{https://orcid.org/0000-0003-3890-0242}{\textcolor[rgb]{0.00,1.00,0.00}{\sf ID}}}$}
\email[]{phycui@nju.edu.cn}
\affiliation{School of Physics, Nanjing University, Nanjing, Jiangsu 210093, China}
\affiliation{Institute for Nonperturbative Physics, Nanjing University, Nanjing, Jiangsu 210093, China}
\author{Craig D.~Roberts%
        $\,^{\href{https://orcid.org/0000-0002-2937-1361}{\textcolor[rgb]{0.00,1.00,0.00}{\sf ID}}}$}
\email[]{cdroberts@nju.edu.cn}
\affiliation{School of Physics, Nanjing University, Nanjing, Jiangsu 210093, China}
\affiliation{Institute for Nonperturbative Physics, Nanjing University, Nanjing, Jiangsu 210093, China}

\begin{abstract}
Octet baryon axial, induced pseudoscalar, and pseudoscalar form factors are computed using a symmetry-preserving treatment of a vector$\,\times\,$vector contact interaction (SCI), thereby unifying them with an array of other baryon properties and analogous treatments of semileptonic decays of pseudoscalar mesons.  The baryons are treated as quark--plus--interacting-diquark bound states, whose structure is obtained by solving a Poincar\'e-covariant Faddeev equation.  The approach is marked by algebraic simplicity, involves no free parameters, and since it is symmetry preserving, all consequences of partial conservation of the axial current are manifest.
It is found that SCI results are consistent with only small violations of SU$(3)$-flavour symmetry, an outcome which may be understood as a dynamical consequence of emergent hadron mass.
The spin-flavour structure of the Poincar\'e-covariant baryon wave functions is expressed in the presence of both flavour-antitriplet scalar diquarks and flavour-sextet axialvector diquarks and plays a key role in determining all form factors.
%
Considering neutral axial currents, SCI predictions for the flavour separation of octet baryon axial charges and, therefrom, values for the associated SU$(3)$ singlet, triplet, and octet axial charges are obtained.  The results indicate that at the hadron scale, $\zeta_{\cal H}$, valence degrees-of-freedom carry roughly 50\% of an octet baryon's total spin.  Since there are no other degrees-of-freedom at $\zeta_{\cal H}$, the remainder may be associated with quark+diquark orbital angular momentum.
\end{abstract}

\maketitle

\end{CJK}


%
\section{Introduction}
The proton $(p)$ is the only stable hadron.  It is the best known bound state in the baryon octet, every other member of which decays.  In many respects, the semileptonic decays of these baryons are the simplest to understand theoretically because the initial and final states involve only one strongly interacting particle.  The archetypal process is neutron $(n)$ $\beta$-decay: $n \to p  e^- \bar\nu_e$, the study of which has a long history \cite{Pauli:1930pc, Fermi:1934hr}.  Notwithstanding that, kindred decays of hyperons have also attracted much attention \cite{Gaillard:1984ny, Cabibbo:2003cu}, in part because they additionally enable access to the Cabibbo-Kobayashi-Maskawa (CKM) matrix element $|V_{us}|$ and thereby complement that provided by $K_{\ell 3}$ decays \cite[Sec.\,12.2.2]{Zyla:2020zbs}.

Within the Standard Model, the semileptonic decay $B \to B^\prime   \ell^-   \nu_\ell$, where
$B$, $B^\prime$ are octet baryons and $\ell$ denotes a lepton, involves a valence-quark $g$ in $B$ transforming into a valence-quark $f$ in $B^\prime$.
Two Poincar\'e-invariant form factors are required to describe the associated axialvector transition matrix element:
{\allowdisplaybreaks
\begin{subequations}
\label{jaxdq0}
\begin{align}
\label{jaxdq}
& J^{B^\prime B}_{5\mu}(K,Q)
:= \langle B^\prime(P^\prime)|{\mathpzc A}^{fg}_{5\mu}(0)|B(P)\rangle \\
\label{jaxdqb}
& =\bar{u}_{B^\prime}(P^\prime) 
\gamma_5  \bigg[ \gamma_\mu G_A^{B^\prime B}(Q^2) +\frac{iQ_\mu}{2M_{B^\prime B}}G_P^{B^\prime B}(Q^2) \bigg]\,u_B(P)\,. 
\end{align}
\end{subequations}
Here
$G_A^{B^\prime B}(Q^2)$ is the axial form factor and $G_P^{B^\prime B}(Q^2)$ is the induced pseudoscalar form factor;
%
%
$P$ and $P^\prime$ are, respectively, the momenta of the initial- and final-state baryons, defined such that the on-shell conditions are fulfilled, $P^{(\prime)}\cdot P^{(\prime)}=-m_{B,B^\prime}^2$, with $m_{B,B^\prime}$ being the baryon masses (we work in Euclidean metric);
$2M_{B^\prime B} = m_{B^\prime}+m_B$;
and $u_{B,B^\prime}(P)$ are the associated Euclidean spinors.  (We have suppressed the spin label. See Ref.\,\cite[Appendix\,B]{Segovia:2014aza} for details.)
Furthermore, $K=(P+P^\prime)/2$ is the average momentum of the system and $Q=P^\prime-P$ the transferred momentum between initial and final states:
\begin{subequations}
\begin{align}
 -K^2  = \tfrac{1}{2}(m_{B^\prime}^2+m_B^2) + \tfrac{1}{4}Q^2 & =: \tfrac{1}{2}\Sigma_{B^\prime B}+\tfrac{1}{4}Q^2 \,, \\
 -K\cdot Q  =\tfrac{1}{2}(m_{B^\prime}^2-m_B^2) & =: \tfrac{1}{2}\Delta_{B^\prime B}\,.
\end{align}
\end{subequations}
}

Hereafter, we consider the isospin symmetry limit $m_u=m_d=:m_l$, \emph{i.e}., degenerate light-quarks, and treat the $s$ valence quark as roughly twenty-times more massive \cite{Zyla:2020zbs}, \emph{viz}.\ $m_s \approx 20\, m_l$.
The general flavour structure is described by the Gell-Mann matrices $\{\lambda^j|j=1,\ldots,8\}$ so that the flavour-nonsinglet axial current operator can be written
\begin{equation}
\label{jaxx}
{\mathpzc A}_{5\mu}^{fg}(x) = \bar{\mathpzc q}(x) {\cal T}^{fg} \gamma_5 \gamma_\mu {\mathpzc q}(x)\,,
\end{equation}
where ${\mathpzc q} = {\rm column}[u,d,s]$
and ${\cal T}^{fg}$ is the valence-quark flavour transition matrix.  Hence, \emph{e.g}., the $s\to u$ transition is described by ${\cal T}^{us} = (\lambda^4+i\lambda^5)/2$.

A related form factor, $G_5^{B^\prime B}(Q^2)$, is associated with an analogous pseudoscalar current
\begin{subequations}
\label{jpsdq0}  
\begin{align}
\label{jpsdq}
 J^{B^\prime B}_{5}(K,Q) &:=
\langle B^\prime(P^\prime)|{\mathpzc P}^{fg}_5(0)|B(P)\rangle \\
&=\bar{u}_{B^\prime}(P^\prime) 
\gamma_5\,G_5(Q^2)\,u_B(P)\,,
\label{G5FF}
\end{align}
\end{subequations}
where ${\mathpzc P}^{fg}_5(x) = \bar{\mathpzc q}(x){\cal T}^{fg}\gamma_5 {\mathpzc q}(x)$
is the flavour-nonsinglet pseudoscalar current operator.  This form factor is important because, amongst other things, owing to dynamical chiral symmetry breaking (DCSB), a corollary of emergent hadron mass (EHM) \cite{Roberts:2020udq, Roberts:2020hiw, Roberts:2021xnz, Roberts:2021nhw, Binosi:2022djx, Papavassiliou:2022wrb}, one has a partial conservation of the axial current (PCAC) relation for each baryon transition $(2{\mathpzc m}_{fg}=m_f+m_g)$:
\begin{subequations}
\label{PCAC}
\begin{align}
& 0 = Q_\mu J^{B^\prime B}_{5\mu}(K,Q) + 2i{\mathpzc m}_{fg}  J^{B^\prime B}_{5}(K,Q) \\
\Rightarrow &\;
G_A^{B^\prime B}(Q^2) - \frac{Q^2}{4 M_{B^\prime B}^2} G_P^{B^\prime B}(Q^2) = \frac{{\mathpzc m}_{fg}}{M_{B^\prime B}} G_5^{B^\prime B}(Q^2) \,.
\end{align}
\end{subequations}
\emph{N.B}.\ The product ${\mathpzc m}_{fg} G_5^{B^\prime B}(Q^2)$ is renormalisation point invariant, not either of these two factors alone.

The identities in Eqs.\,\eqref{PCAC} are valid for all $Q^2$.
They state that the longitudinal part of the axialvector current is completely determined by the kindred pseudoscalar form factor and possesses a strength modulated by the ratio of the sum of current-quark masses involved in the transition divided by the sum of the masses of the baryons involved.
The former are determined by Higgs boson couplings into quantum chromodynamics (QCD), whereas the latter are largely determined by the scale of EHM.  Hence, this $Q$-divergence is a measure of the interplay between Nature's two known mass generating mechanisms.

Specialising to the case of neutron $\beta$ decay, Eqs.\,\eqref{PCAC} entail the well-known Goldberger-Treiman relation and ensure reliability of the pion pole dominance approximation for $G_P^{pn}$.
Considering instead a prominent hyperon decay, \emph{e.g}., $\Lambda \to p e^- \bar\nu_e$, one recognises that
$G_5^{p \Lambda}$ has a pole at the charged kaon mass, \emph{i.e}., when $Q^2+m_K^2=0$.
Since $G_A^{p \Lambda}$ is tied to the transverse part of the axial current, so regular in the neighbourhood of $m_K^2$, then $G_P^{p \Lambda}$ also has a pole at $m_K$.
Further, defining a $Kp\Lambda$ form factor as follows:
\begin{equation}
G_5^{p \Lambda}(Q^2) =: \frac{m_K^2}{Q^2+m_K^2} \frac{2 f_K}{m_u+m_s}G_{Kp\Lambda}(Q^2)\,,
\label{CouplingKpLambda}
\end{equation}
where $f_K$ is the kaon leptonic decay constant, then Eqs.\,\eqref{PCAC} entail
\begin{equation}
G_A^{p\Lambda}(0) = \frac{2 f_K}{m_p+m_\Lambda}G_{Kp\Lambda}(0)\,,
\end{equation}
providing an estimate of the $Kp\Lambda$ coupling in terms of the axialvector $\Lambda\to p$ transition form factor at the maximum recoil point.  As we shall see in our analysis, this relation is accurate to better than 1\%.

The evidently diverse physics relevance of octet baryon axialvector transitions highlights the importance of calculating the associated form factors.  However, despite their being some of the simplest baryonic processes to consider, this does not mean their calculation is simple.  Studies of meson semileptonic transitions \cite{Chen:2012txa, Xu:2021iwv, Yao:2021pyf, Yao:2021pdy, Xing:2022sor} have revealed that delivering predictions for the required processes demands reliable calculations of the Poincar\'e-covariant hadron wave functions and the related axialvector interaction currents and careful symmetry-preserving treatments of the matrix elements involved.

Given their role in understanding modern neutrino experiments \cite{Mosel:2016cwa, Alvarez-Ruso:2017oui, Hill:2017wgb, Gysbers:2019uyb, Lovato:2020kba}, the nucleon axial and pseudoscalar form factors have recently been the focus of many studies, using continuum and lattice methods, \emph{e.g}., Refs.\,\cite{Anikin:2016teg, Chen:2020wuq, Chen:2021guo, ChenChen:2022qpy, Alexandrou:2017hac, Jang:2019vkm}.
Regarding hyperon semileptonic decays, analyses using an array of tools may be found, \emph{e.g}.,
in Refs.\,\cite{Faessler:2008ix, Ledwig:2014rfa, Yang:2015era, Ramalho:2015jem, Yang:2018idi, Qi:2022sus,
Erkol:2009ev, Green:2017keo, Bali:2022qja}.
Herein, we employ continuum Schwinger function methods (CSMs) \cite{Eichmann:2016yit, Burkert:2017djo, Qin:2020rad} to complement this body of work on octet baryon axialvector transitions.
Namely, we construct approximations to the transition matrix elements using solutions to a symmetry-preserving collection of integral equations for the relevant $n$-point Schwinger functions, $n=2-6$.  This is now possible following development of a realistic axial current for baryons \cite{Chen:2020wuq, Chen:2021guo}.

One could extend to hyperons the QCD-kindred framework used in Refs.\,\cite{Chen:2020wuq, Chen:2021guo, ChenChen:2022qpy} to compute all form factors associated with the nucleon axial and pseudoscalar currents.  However, that would require significant effort.  An expeditious alternative is to simplify the analysis by using the symmetry-preserving formulation of a vector$\,\times\,$vector contact interaction (SCI) introduced in Refs.\,\cite{GutierrezGuerrero:2010md, Roberts:2011wy, Chen:2012qr}.  In so doing, one ensures algebraic simplicity and, very importantly, provides for the parameter-free unification of octet baryon axialvector transitions with an array of other baryon properties \cite{Wilson:2011aa, Xu:2015kta, Yin:2021uom, Raya:2021pyr} and studies of the semileptonic decays of pseudoscalar mesons \cite{Chen:2012txa, Xu:2021iwv, Xing:2022sor}.
By choosing this approach, we are profiting from numerous studies \cite{Wang:2013wk, Xu:2015kta, Bedolla:2015mpa, Bedolla:2016yxq, Serna:2017nlr, Raya:2017ggu, Zhang:2020ecj, Yin:2019bxe, Yin:2021uom, Raya:2021pyr, Lu:2021sgg, Gutierrez-Guerrero:2021rsx} which have revealed that, when interpreted judiciously, SCI predictions provide a valuable quantitative guide.  In fact, SCI results typically deliver both a useful first estimate of a given observable and a means of checking the validity of algorithms employed in calculations that rely (heavily) upon high performance computing.

In Sec.\,\ref{SecTwo}, we sketch the Faddeev equation used to describe baryons as quark--plus--interacting-diquark bound states and the current which guarantees preservation of all PCAC identities.
The description is complemented by an extensive appendix, which provides a detailed explanation of the SCI and its results for every element that appears in the Faddeev equations and currents.
Using that information, Sec.\,\ref{SecThree} presents and analyses SCI predictions for octet baryon axial, induced pseudoscalar, and pseudoscalar transition form factors.
This is followed in Sec.\,\ref{SecFour} with a discussion of the flavour separation of  octet baryon axial charges and their relation to the fraction of baryon spin carried by valence degrees of freedom.
Section~\ref{epilogue} is a summary and perspective.

\begin{figure}[t]
\centerline{%
\includegraphics[clip, height=0.14\textwidth, width=0.45\textwidth]{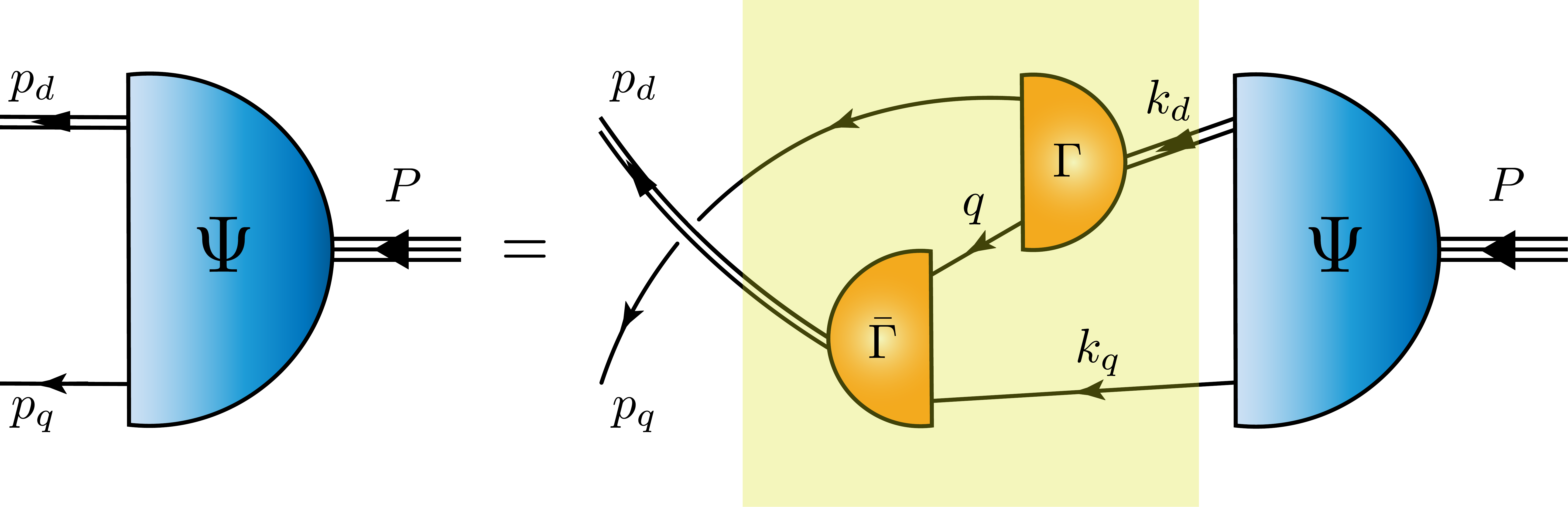}}
\caption{\label{figFaddeev}
Integral equation for the Poincar\'e-covariant matrix-valued function $\Psi$, the Faddeev amplitude for a baryon with total momentum $P=p_q+p_d=k_q+k_d$ constituted from three valence quarks, two of which are always contained in a nonpointlike, interacting diquark correlation. $\Psi$ describes the relative momentum correlation between the dressed-quarks and -diquarks.
Legend. \emph{Shaded rectangle} -- Faddeev kernel;
\emph{single line} -- dressed-quark propagator [Appendix\,\ref{AppendixSCI}];
$\Gamma$ -- diquark correlation amplitude and \emph{double line} -- diquark propagator [Appendix\,\ref{Appendixdiquarks}].
Ground-state $J=1/2^+$ baryons contain both flavour-antitriplet--scalar and flavour-sextet--axialvector diquarks [Appendix\,\ref{AppendixFaddeev}].}
\end{figure}

\section{Baryons and their axial current}
\label{SecTwo}
\begin{figure}[!t]
\centerline{\includegraphics[clip, height=0.33\textwidth, width=0.47\textwidth]{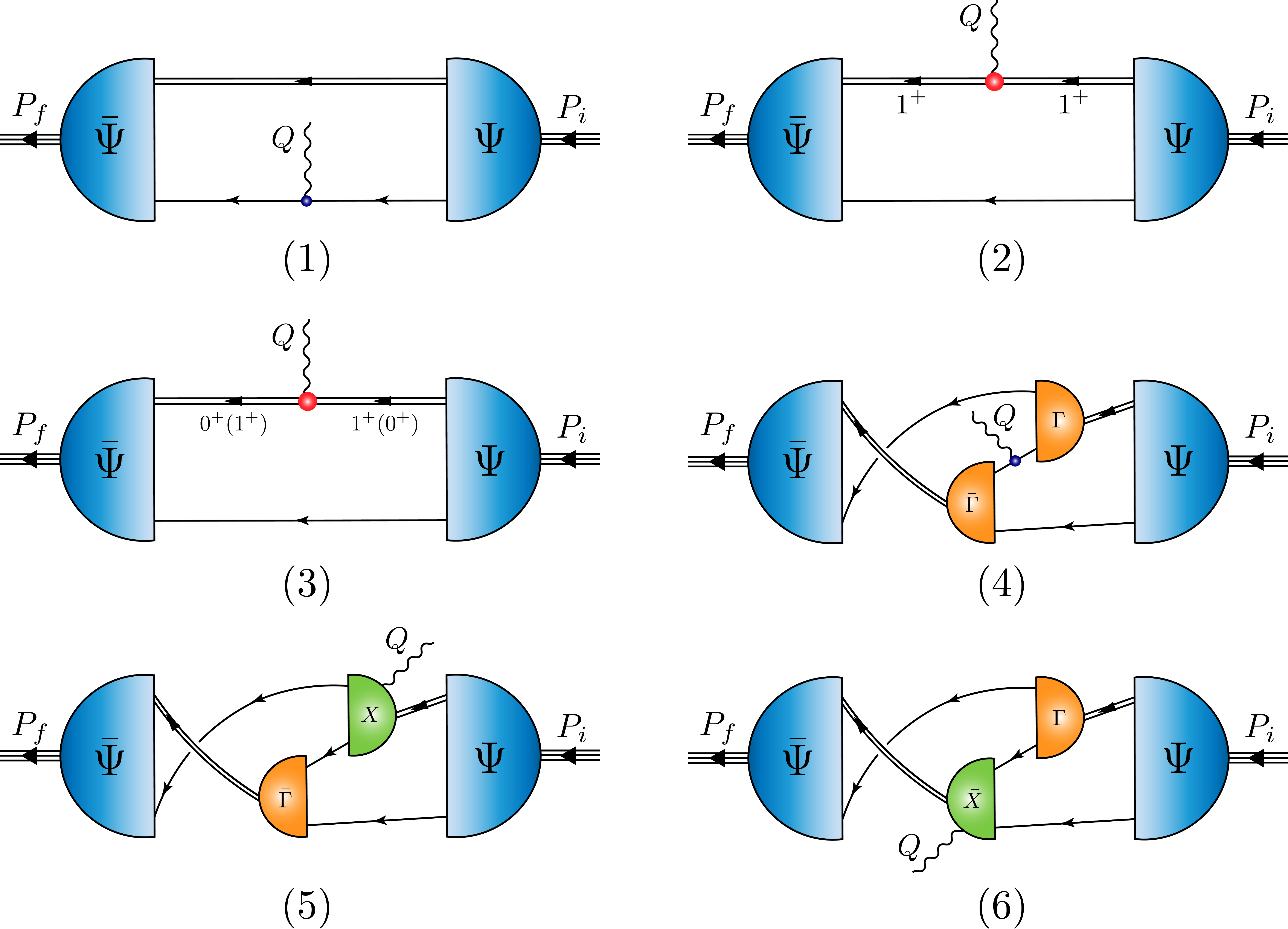}}
\caption{\label{figcurrent} Currents that ensure PCAC for on-shell baryons which are described by the Faddeev amplitudes produced by the equation depicted in Fig.\,\ref{figFaddeev}: \emph{single line}, dressed-quark propagator; \emph{undulating line}, axial or pseudoscalar current; $\Gamma$,  diquark correlation amplitude; \emph{double line}, diquark propagator; and $\chi$, seagull terms.
A legend is provided in Table~\ref{DiagramLegend} with details in Appendix\,\ref{AppendixCurrents}.
}
\end{figure}

Our analyses of octet baryon axialvector transition form factors rest on solutions of the Poincar\'e-covariant Faddeev equation depicted in Fig.\,\ref{figFaddeev}, which, when inserted into the diagrams drawn in Fig.\,\ref{figcurrent}, deliver a result for the current in Eq.\,\eqref{jaxdq0} that ensures Eqs.\,\eqref{PCAC} and all their corollaries for each transition.  Details are presented in Refs.\,\cite{Chen:2020wuq, Chen:2021guo}.  For subsequent use, in Table~\ref{DiagramLegend} we identify a useful separation of the current in Fig.\,\ref{figcurrent}.

Evidently, we have adopted the quark--plus--interacting-diquark picture of baryon structure introduced in Refs.\,\cite{Cahill:1988dx, Reinhardt:1989rw, Efimov:1990uz}, of which an updated perspective is provided in Refs.\,\cite{Barabanov:2020jvn, Lu:2022cjx, Liu:2022ndb, Eichmann:2022zxn}.  In this approach, there are two contributions to binding within a baryon \cite{Segovia:2015ufa}.  One part is expressed in the formation of tight (but not pointlike) quark+quark correlations. It is augmented by the attraction generated by the quark exchange depicted in the shaded area of Fig.\,\ref{figFaddeev}, which ensures that diquark correlations within the baryon are fully dynamical.  Namely, no quark is special because each one participates in all diquarks to the fullest extent allowed by its quantum numbers. The continual rearrangement of the quarks guarantees, \emph{inter alia}, that the baryon's dressed-quark wave function complies with Pauli statistics.  The spin-flavour wave function of $J^P=1/2^+$ ground-state baryons is overwhelmingly dominated by flavour-antitriplet--scalar and flavour-sextet--axialvector diquarks \cite{Chen:2012qr, Chen:2017pse, Yin:2019bxe, Yin:2021uom, Gutierrez-Guerrero:2021rsx, Raya:2021pyr, Eichmann:2022zxn}.

\begin{table}[b!]
\caption{\label{DiagramLegend}
Enumeration of terms in the current drawn in Fig.\,\ref{figcurrent}.}
\begin{enumerate}
\item Diagram~1, two distinct terms: $\langle J \rangle^{S}_{\rm q}$ -- probe strikes dressed-quark with scalar diquark spectator; and $\langle J \rangle^{A}_{\rm q}$ -- probe strikes dressed-quark with axialvector diquark spectator.
\item Diagram~2: $\langle J \rangle^{AA}_{\rm qq}$ -- probe strikes axialvector diquark with dressed-quark spectator.
\item Diagram~3: $\langle J \rangle^{\{SA\}}_{\rm qq}$ -- probe mediates transition between scalar and axialvector diquarks, with dressed-quark spectator.
\item Diagram~4, three terms:
    $\langle J \rangle_{\rm ex}^{SS}$ -- probe strikes dressed-quark ``in-flight'' between one scalar diquark correlation and another;
    $\langle J \rangle_{\rm ex}^{\{SA\}}$ -- dressed-quark ``in-flight'' between a scalar diquark correlation and an axialvector correlation;
    and $\langle J \rangle_{\rm ex}^{AA}$ -- ``in-flight'' between one axialvector correlation and another.
\item Diagrams~5 and 6 -- seagull diagrams describing the probe coupling into the diquark correlation amplitudes: $\langle J\rangle_{\rm sg}$.  There is one contribution from each diagram to match every term in Diagram (4).
\end{enumerate}
\end{table}

The first step in our analysis of octet baryon transitions is the SCI calculation of every line, amplitude and vertex in Figs.\,\ref{figFaddeev}, \ref{figcurrent}.  These calculations are described in Appendix~\ref{AppendixSupplement}.  Combining the results and using sensibly chosen projection operators, one readily arrives at predictions for the baryon axial and pseudoscalar form factors in Eqs.\,\eqref{jaxdqb}, \eqref{G5FF}.  \emph{N.B}.\ Eq.\,\eqref{jaxdq} entails that $G_A^{B^\prime B}$ is entirely determined by the $Q$-transverse part of the baryon axial current \cite{Chen:2021guo}.

\section{Calculated Form Factors}
\label{SecThree}
\subsection{Axial}
\label{GAsection}
In the isospin-symmetry limit, there are six distinct charged current semileptonic transitions between octet baryons.  We record our predictions for the associated $G_A(Q^2=0)$ values in Table~\ref{GAzero}.  In the Cabibbo model of such transitions, which assumes SU$(3)$-flavour symmetry, the couplings in Table~\ref{GAzero} are described by just two distinct parameters \cite[Table~1]{Cabibbo:2003cu}: $D$, $F$.  In these terms, a least-squares fit to the SCI results produces
\begin{equation}
\label{DFvalues}
D = 0.78\,, \; F = 0.43\,,\; F/D = 0.56\,,
\end{equation}
with a mean absolute relative error between SCI results and Cabibbo fit of just 3(2)\%.  Evidently, confirming the conclusion of many studies, the SCI predicts that the violation of SU$(3)$ symmetry in these transitions is small.  This is also manifest in the comparison between $n\to p$ and $\Xi^0\to \Sigma^+$.  The former is a $d\to u$ transition, the latter is $s\to u$; yet, in the Cabibbo model, $G_A^{\Sigma^+\Xi^0}(0)=G_A^{pn}(0)$, and this identity is accurate to 4\% in the SCI calculation.  Likewise in experiment.

It is worth providing additional context for the results in Eq.\,\eqref{DFvalues}.  We therefore note that
a covariant baryon chiral perturbation theory analysis of semileptonic hyperon decays yields $D=0.80(1)$, $F=0.47(1)$, $F/D=0.59(1)$ \cite{Ledwig:2014rfa};
and a three-degenerate-flavour lattice QCD (lQCD) computation yields $F/D = 0.61(1)$ \cite{Bali:2022qja}.

\begin{table}[t]
\caption{\label{GAzero}
SCI predictions for $g_A^{B^\prime B}=G_A^{B^\prime B}(Q^2=0)$ compared with experiment \cite{Zyla:2020zbs} and other calculations:
Lorentz covariant quark model \cite{Faessler:2008ix};
covariant baryon chiral perturbation theory \cite{Ledwig:2014rfa};
and a lQCD study \cite{Erkol:2009ev}, which used large pion masses ($m_\pi = 0.55\,$-$\,1.15\,$GeV) and quoted error estimates that are primarily statistical.
%
}
\begin{center}
\begin{tabular*}
{\hsize}
{
l@{\extracolsep{0ptplus1fil}}
|c@{\extracolsep{0ptplus1fil}}
|c@{\extracolsep{0ptplus1fil}}
|c@{\extracolsep{0ptplus1fil}}
|c@{\extracolsep{0ptplus1fil}}
|c@{\extracolsep{0ptplus1fil}}
|c@{\extracolsep{0ptplus1fil}}}\hline
%
& $n\to p\ $ & $\Sigma^- \to \Lambda\ $ & $\Lambda \to p\ $ & $\Sigma^- \to n\ $ & $\Xi^0 \to \Sigma^+\ $ & $\Xi^- \to \Lambda\ $ \\\hline
SCI\; & $1.24\phantom{(3)}\ $ & $0.66\phantom{(3)}\ $ & $-0.82\phantom{(2)}\ $ & $0.34\phantom{(2)}\ $ & $1.19\phantom{(5)}\ $ & $0.23\phantom{(6)}\ $\\   \hline
\cite{Zyla:2020zbs}\; & $1.28\phantom{(3)}\ $ & $0.57(3)\ $ & $-0.88(2)\ $ & $0.34(2)\ $ & $1.22(5)\ $ & $0.31(6)\ $\\
\cite{Faessler:2008ix} & $1.27\phantom{(3)}\ $ & $0.63\phantom{(2)}\ $ & $-0.89\phantom{(2)}\ $ & $0.26\phantom{(2)}\ $ & $1.25\phantom{(4)}\ $ & $0.33\phantom{(4)}\ $\\
\cite{Ledwig:2014rfa} & $1.27\phantom{(3)}\ $ & $0.60(2)\ $ & $-0.88(2)\ $ & $0.33(2)\ $ & $1.22(4)\ $ & $0.21(4)\ $\\
\cite{Erkol:2009ev} & $1.31(2)\ $ & $0.66(1)\ $ & $-0.95(2)\ $ & $0.34(1)\ $ & $1.28(3)\ $ & $0.27(1)\ $\\
\hline
\end{tabular*}
\end{center}
\end{table}

In considering the empirical fact of approximate SU$(3)$-flavour symmetry in the values of octet baryon axial transition charges, one should note that it is not a direct consequence of any basic symmetry.  Hence, the apparent near-symmetry is actually a dynamical outcome.
The underlying source of any SU$(3)$-flavour symmetry breaking is the Higgs-boson generated splitting between the current masses of the $s$ and $l=u,d$ valence quarks.  However, as noted above, $m_s/m_l \approx 20$.  Therefore, something must be strongly suppressing the expression of this difference in observable quantities.

The responsible agent is EHM \cite{Roberts:2020udq, Roberts:2020hiw, Roberts:2021xnz, Roberts:2021nhw, Binosi:2022djx, Papavassiliou:2022wrb}.
For example, leptonic weak decays of pseudoscalar mesons proceed via the axial current and $f_K/f_\pi \approx 1.2$.  These decay constants are order parameters for chiral symmetry breaking and that effect is predominantly dynamical for Nature's three lighter quarks \cite[Fig.\,2.5]{Roberts:2021nhw}.
Similarly, considering the axial form factors for semileptonic decays of heavy+light pseudoscalar mesons to light vector meson final states, one finds SU$(3)$-flavour symmetry breaking on the order of 10\% \cite{Xing:2022sor}.
Finally, comparing the hadron-scale valence-quark distribution functions of the kaon and pion, one learns that the $u$ quark carries 6\% less of the kaon's light-front momentum than does the $u$-quark in the pion \cite{Cui:2020dlm, Cui:2020tdf}.

Focusing on the case in hand, \emph{i.e}., octet baryon semileptonic transitions, $m_s/m_l \approx 20$ leads to a dressed-quark mass-ratio $M_s/M_l \approx 1.4$ -- Table~\ref{Tab:DressedQuarks}; namely, a huge suppression owing to EHM.
In turn, this is expressed as a $\sim 14$\% difference in diquark masses,
smaller differences in diquark correlation amplitudes,
and, consequently, differences of even smaller magnitude ($\sim 3$\%) between the leading scalar-diquark components of the Faddeev amplitudes of the baryons involved.
In addition, Tables~\ref{interpolatorcoefficientsdu}, \ref{interpolatorcoefficientsus} reveal that the $s\to u$ and $d\to u$ quark-level weak transitions are similar in strength -- unsurprising given that these axial vertices are obtained by solving Bethe-Salpeter equations akin to those that yield the diquark correlation amplitudes.
Finally, therefore, regarding the $n\to p$\,:\,$\Xi^0\to \Sigma^+$ comparison, \emph{e.g}., Table~\ref{GA0diagrams} reveals that the scalar diquark components dominate the transition; hence, these transitions should have similar strengths.

\begin{table}[t]
\caption{\label{GA0diagrams}
Diagram separation of octet baryon axial transition charges, presented as a fraction of the total listed in Table~\ref{GAzero}\,--\,Row~1 and made with reference to Fig.\,\ref{figcurrent}.
}
\begin{center}
\begin{tabular*}
{\hsize}
{
l@{\extracolsep{0ptplus1fil}}
|l@{\extracolsep{0ptplus1fil}}
l@{\extracolsep{0ptplus1fil}}
l@{\extracolsep{0ptplus1fil}}
l@{\extracolsep{0ptplus1fil}}
l@{\extracolsep{0ptplus1fil}}
l@{\extracolsep{0ptplus1fil}}
l@{\extracolsep{0ptplus1fil}}}\hline
 & $\langle J \rangle^{S}_{\rm q}$  & $\langle J \rangle^{A}_{\rm q}$ &$\langle J \rangle^{AA}_{\rm qq}$ & $\langle J \rangle^{\{SA\}}_{\rm qq}$
 & $\langle J \rangle_{\rm ex}^{SS}$
 & $\langle J \rangle_{\rm ex}^{\{SA\}}$
 & $\langle J \rangle_{\rm ex}^{AA}$  \\\hline
$g_A^{pn}$ & $0.29\ $ & $\phantom{-}0.013\ $ & $\phantom{-}0.072\ $ & $0.35\ $ & $\phantom{-}0.19\ $ & $\phantom{-}0.051\ $& $0.028\ $\\
$g_A^{\Sigma^- \Lambda}$ & $0.27\ $ & $\phantom{-}0.016\ $ & $\phantom{-}0.023\ $ & $0.42\ $ & $\phantom{-}0.28\ $ & $-0.008\ $ & \\
$g_A^{p \Lambda}$ & $0.45\ $ & & $\phantom{-}0.083\ $ & $0.33\ $ & $\phantom{-}0.082\ $ & $\phantom{-}0.044\ $& $0.013\ $\\
$g_A^{n\Sigma^-}$ & & $\phantom{-}0.13\ $ & $-0.051\ $ & $0.57\ $ & $\phantom{-}0.42\ $ & $-0.076\ $& $0.008\ $\\
$g_A^{\Sigma^+ \Xi^0}$ & $0.41\ $ & $\phantom{-}0.011\ $ & $\phantom{-}0.064\ $ & $0.36\ $ & $\phantom{-}0.12\ $ & $\phantom{-}0.020\ $& $0.013\ $\\
$g_A^{\Lambda \Xi^-}$ & $1.02\ $ & $-0.072\ $ & $\phantom{-}0.12\ $ & $0.12\ $ & $-0.28\ $ & $\phantom{-}0.023\ $ & $0.076\ $\\
\hline
\end{tabular*}
\end{center}
\end{table}

Table~\ref{GA0diagrams} highlights a curious feature of the quark+diquark picture; namely, the $s\to u$ quark $\Sigma^-\to n$ transition receives no contribution from Diagram~1 in Fig.\,\ref{figcurrent} because the only scalar diquark component in $\Sigma^-$ is $d[ds]$ and the neutron contains no $[ds]$ diquark.  Nevertheless, scalar diquarks are still dominant contributors to $g_A^{n\Sigma^-}$ via Diagrams~3 and 4.  It is also worth recalling that since axial form factors derive solely from $Q$-transverse pieces of the baryon current \cite{Chen:2021guo}, there are no seagull contributions to $G_A^{B^\prime B}$.

Notwithstanding the dominance of scalar diquark components, Table~\ref{GA0diagrams} reveals that axialvector correlations also play a material role in the transitions.  For instance, $\langle J \rangle_{\rm qq}^{SA}$ is large in all cases; yet, would vanish if axialvector diquarks were ignored in forming the picture of baryon structure.  Their impact is further highlighted below.

\begin{figure}[t]
\vspace*{2ex}

\leftline{\hspace*{0.5em}{\large{\textsf{A}}}}
\vspace*{-4ex}
\includegraphics[width=0.42\textwidth]{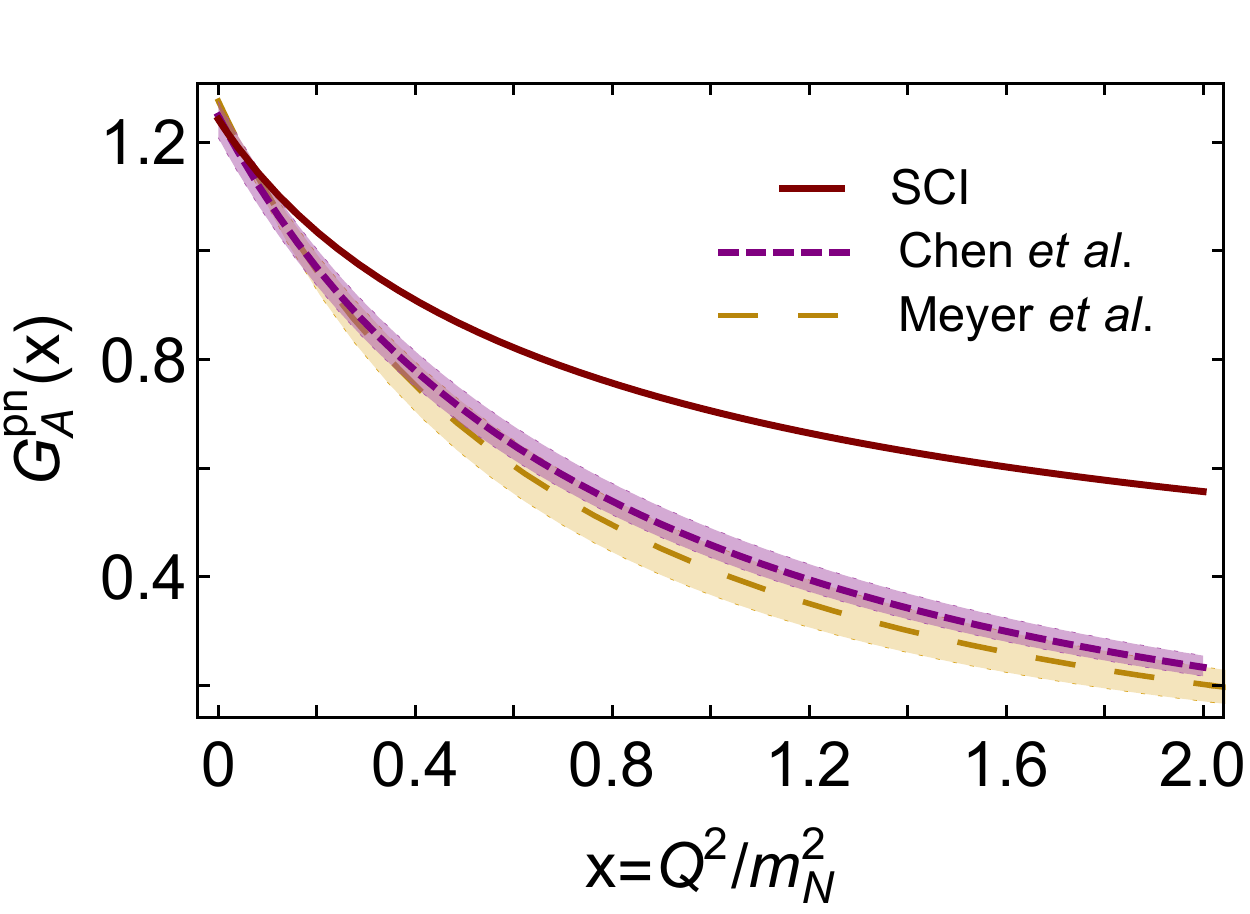}
\vspace*{1ex}

\leftline{\hspace*{0.5em}{\large{\textsf{B}}}}
\vspace*{-4ex}
\includegraphics[width=0.425\textwidth]{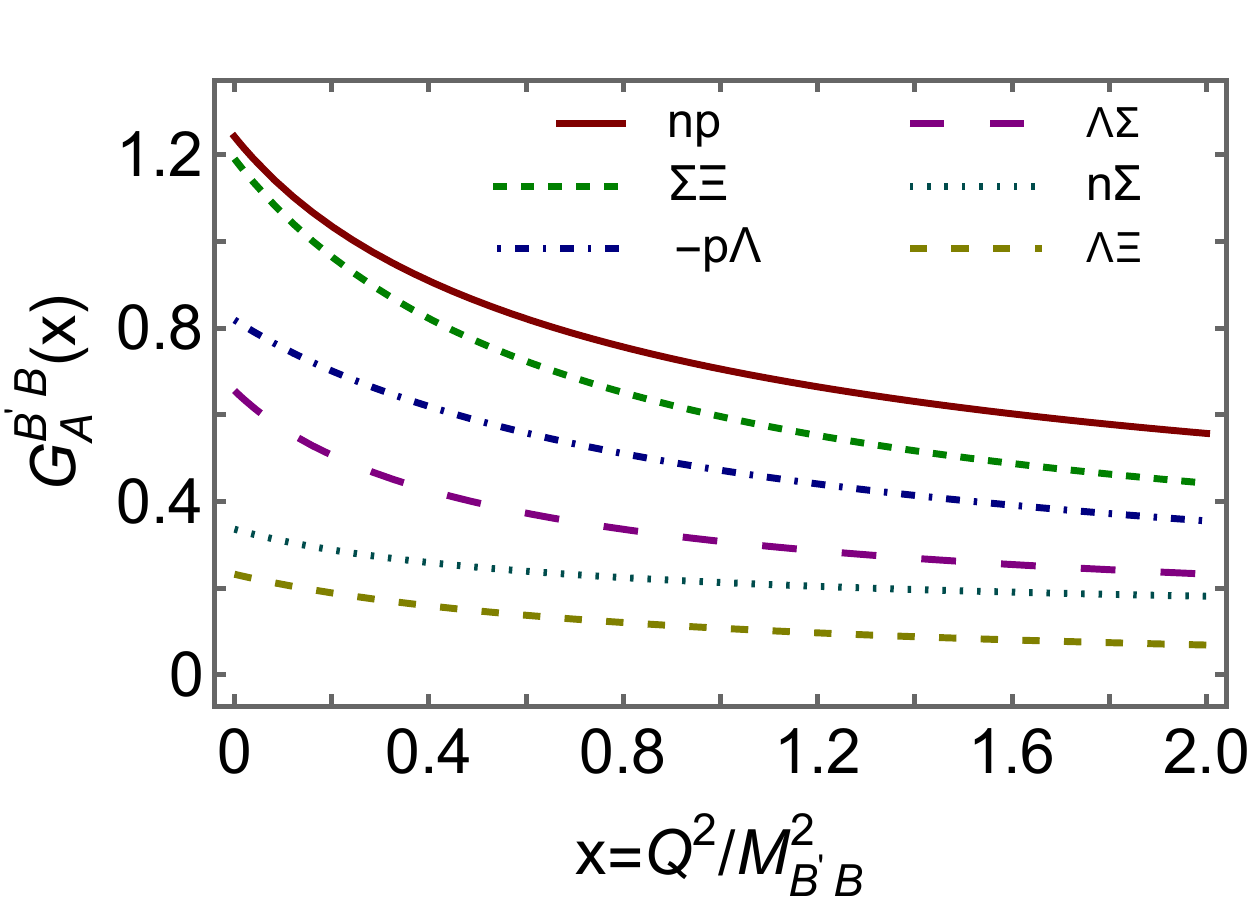}
\vspace*{1ex}

\leftline{\hspace*{0.5em}{\large{\textsf{C}}}}
\vspace*{-4ex}
\includegraphics[width=0.425\textwidth]{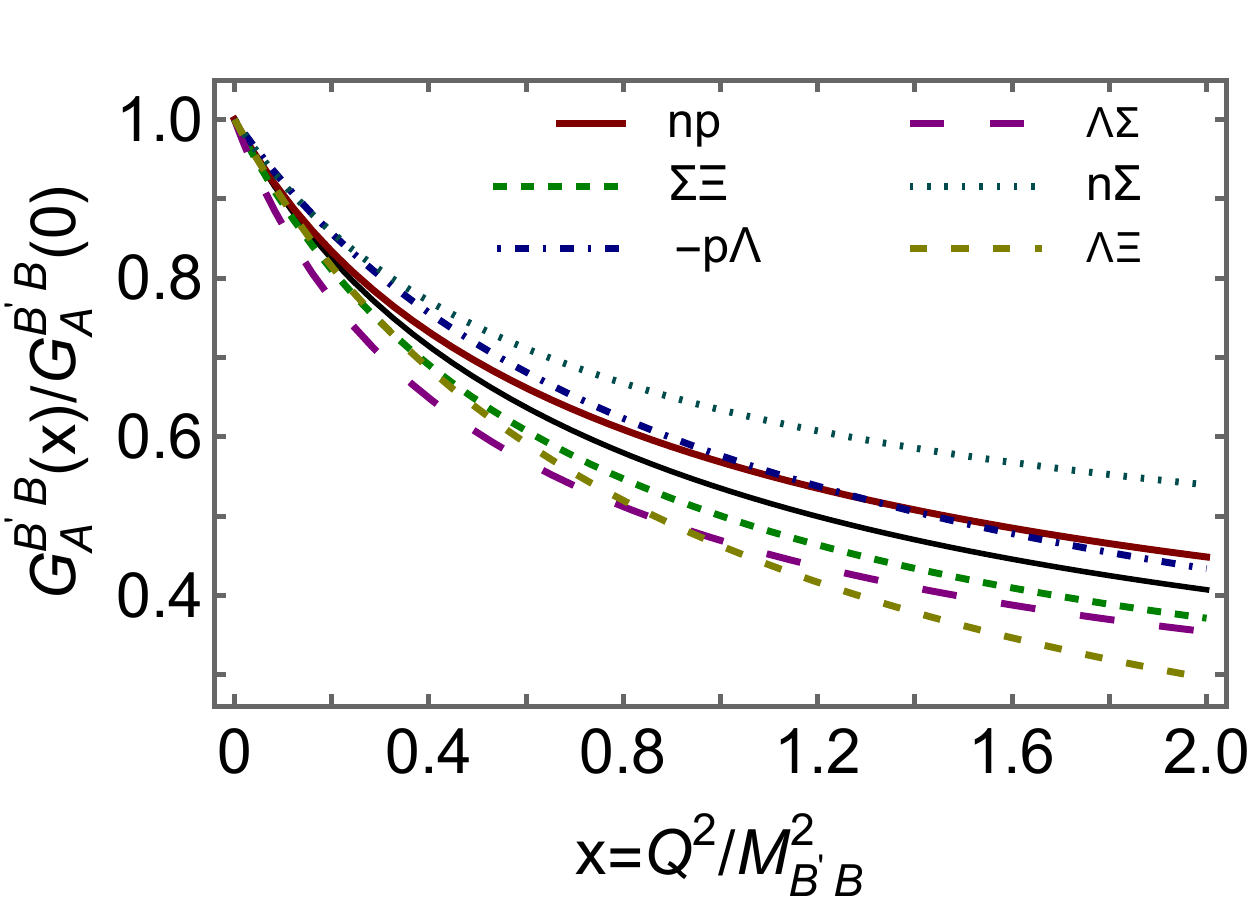}
\caption{\label{FigGAx}
{\sf Panel A}.
$G_A^{pn}(x=Q^2/m_N^2)$: SCI result computed herein -- solid red curve; prediction from Ref.\,\cite{Chen:2021guo} -- short-dashed purple curve within like-coloured band; and dipole fit to data \cite{Meyer:2016oeg} -- long-dashed gold curve within like-coloured band.
{\sf Panel B}.  Complete array of SCI predictions for octet baryon axial transition form factors: $G_A^{B^\prime B}(x=Q^2/M_{B^\prime B}^2)$
{\sf Panel C}.  As in Panel B, but with each form factor normalised to unity at $x=0$.  The thinner solid black curve is a pointwise average of the six transition form factors.
}
\end{figure}

Our calculated SCI result for $G_A^{pn}(Q^2= x m_N^2)$ is reliably interpolated using the function in Eq.\,\eqref{SCIinterpolationsGA} with the coefficients in Table~\ref{InterpolationsGA}A.  It is drawn in Fig.\,\ref{FigGAx}A and compared with both the CSM prediction from Ref.\,\cite{Chen:2021guo}, produced with QCD-like momentum dependence for all elements in Figs.\,\ref{figFaddeev}, \ref{figcurrent}, and a dipole fit to low-$Q^2$ data \cite{Meyer:2016oeg}.
As usual with SCI predictions: the $x \lesssim M_l^2$ results are quantitatively sound ($M_l$ is the dressed-mass of the lighter quarks -- Table~\ref{Tab:DressedQuarks}); but form factor evolution with increasing $x$ is too slow \cite{GutierrezGuerrero:2010md, Roberts:2011wy, Chen:2012qr}, \emph{i.e}., SCI form factors are too hard at spacelike momenta.

The complete array of ground-state octet baryon axial transition form factors is plotted in Fig.\,\ref{FigGAx}B.  Interpolations of these functions are obtained using Eq.\,\eqref{SCIinterpolationsGA} and the appropriate coefficients from Table~\ref{InterpolationsGA}A.

Fig.\,\ref{FigGAx}C presents the curves in Fig.\,\ref{FigGAx}B renormalised to unity at $x=0$ along with the pointwise average of the renormalised functions.
%
Introducing a dimensionless radius-squared associated with the curves drawn, \emph{viz}.
\begin{equation}
(\hat r_A^{B^\prime B})^2 = -6 M_{B^\prime B} \frac{d}{dQ^2}[G_A^{B^\prime B}(Q^2)/G_A^{B^\prime B}(0)],
\end{equation}
in terms of which the usual radius is $r_A^{B^\prime B} = \hat r_A^{B^\prime B}/M_{B^\prime B}$,
one arrives at the following comparisons:
\begin{equation}
\begin{array}{ccccc}
\hat r_A^{\Sigma^- \Lambda}/\hat r_A^{pn} & \hat r_A^{\Lambda p}/\hat r_A^{pn} & \hat r_A^{n \Sigma^-}/\hat r_A^{pn} &
\hat r_A^{\Sigma^+ \Xi^0}/\hat r_A^{pn} & \hat r_A^{\Lambda \Xi^-}/\hat r_A^{pn} \\
1.22 & 0.89 & 0.90 & 1.05 & 1.00
\end{array}\,, \label{radiiratio}
\end{equation}
which quantify the pattern that can be read ``by eye'' from Fig.\,\ref{FigGAx}C.  Evidently, removing the $M_{B^\prime B}$ kinematic factor has revealed a fairly uniform collection of axial transition form factors: the mean value of the ratio in Eq.\,\eqref{radiiratio} is $1.01(13)$.
Given that SCI form factors are typically hard, the individual SCI radii are likely too small; nevertheless, their size relative to $\hat r_A^{pn}$ should be a reliable guide.  So for a physical interpretation of these ratios, we note that comparing the SCI result for $\hat r_A^{pn}$ with that in Ref.\,\cite{Chen:2021guo}, one has $\hat r_{A\,{\rm SCI}}^{pn}/\hat r_{A\,\mbox{\footnotesize\cite{Chen:2021guo}}}^{pn}=0.76$ and $\hat r_{A\,\mbox{\footnotesize\cite{Chen:2021guo}}}^{pn}=3.40(4)$.  The dipole fit to data in Fig.\,\ref{FigGAx}A yields $r_{A\,\mbox{\footnotesize\cite{Meyer:2016oeg}}}^{pn}=3.63(24)$.

Considering the $x$-dependence of the axial transition form factors displayed in Fig.\,\ref{FigGAx}C, it is worth noting that at $x=2$ the mean absolute value of the relative deviation from the average curve is 16(8)\%.
Apparently, the magnitude of SU$(3)$-flavour symmetry breaking increases with $Q^2$, \emph{i.e}., as details of baryon structure are probed with higher precision.  This may also be highlighted by comparing the $x=2$ values of the $n\to p$ and $\Xi^0\to \Sigma^+$ curves in Fig.\,\ref{FigGAx}C: at $x=2$, the ratio is $\approx 1.2$.  It would be unity in the case of SU$(3)$-flavour symmetry.

\begin{table}[t]
\caption{\label{GP0diagrams}
Diagram separated contributions to $Q^2=0$ values of octet baryon induced pseudoscalar transition form factors, $G_P^{B^\prime B}$, presented as a fraction of the total listed in Table~\ref{InterpolationsGA}B\,--\,Column~1 and made with reference to Fig.\,\ref{figcurrent}.
}
\begin{center}
\begin{tabular*}
{\hsize}
{
l@{\extracolsep{0ptplus1fil}}
|c@{\extracolsep{0ptplus1fil}}
c@{\extracolsep{0ptplus1fil}}
c@{\extracolsep{0ptplus1fil}}
c@{\extracolsep{0ptplus1fil}}
c@{\extracolsep{0ptplus1fil}}
c@{\extracolsep{0ptplus1fil}}}\hline
 & $\langle J \rangle^{S}_{\rm q}$  & $\langle J \rangle^{A}_{\rm q}$ &$\langle J \rangle^{AA}_{\rm qq}$ & $\langle J \rangle^{\{SA\}}_{\rm qq}$
 & $\langle J \rangle_{\rm ex}$
 & $\langle J \rangle_{\rm sg}$ \\\hline
$g_P^{pn}$ & $0.54\ $ & $\phantom{-}0.051\ $ & $\phantom{-}0.072\ $ & $0.35\phantom{0}\ $ & $\phantom{-}0.018\ $ & $-0.039\ $ \\
$g_P^{\Sigma^- \Lambda}$ & $0.43\ $ & $\phantom{-}0.054\ $ & $\phantom{-}0.023\ $ & $0.42\phantom{0}\ $ & $\phantom{-}0.023\ $ & $\phantom{-}0.051\ $ \\
$g_P^{p \Lambda}$ & $0.81\ $ & & $\phantom{-}0.073\ $ & $0.32\phantom{0}\ $ & $-0.064\ $ & $-0.14\phantom{0}\ $\\
$g_P^{n\Sigma^-}$ & & $\phantom{-}0.46\phantom{0}\ $ & $-0.047\ $ & $0.56\phantom{0}\ $ & $-0.19\phantom{0}\ $ & $\phantom{-}0.22\phantom{0}\ $ \\
$g_P^{\Sigma^+ \Xi^0}$ & $0.66\ $ & $\phantom{-}0.036\ $ & $\phantom{-}0.061\ $ & $0.33\phantom{0}\ $ & $-0.048\ $ & $-0.033\ $\\
$g_P^{\Lambda \Xi^-}$ & $1.57\ $ & $-0.23\phantom{0}\ $ & $\phantom{-}0.10\phantom{0}\ $ & $0.091\ $ & $\phantom{-}0.13\phantom{0}\ $ & $-0.66\phantom{0}\ $ \\
\hline
\end{tabular*}
\end{center}
\end{table}

\subsection{Induced pseudoscalar}
\label{SecGP}
\begin{figure}[t]
\vspace*{2ex}

\leftline{\hspace*{0.5em}{\large{\textsf{A}}}}
\vspace*{-4ex}
\includegraphics[width=0.42\textwidth]{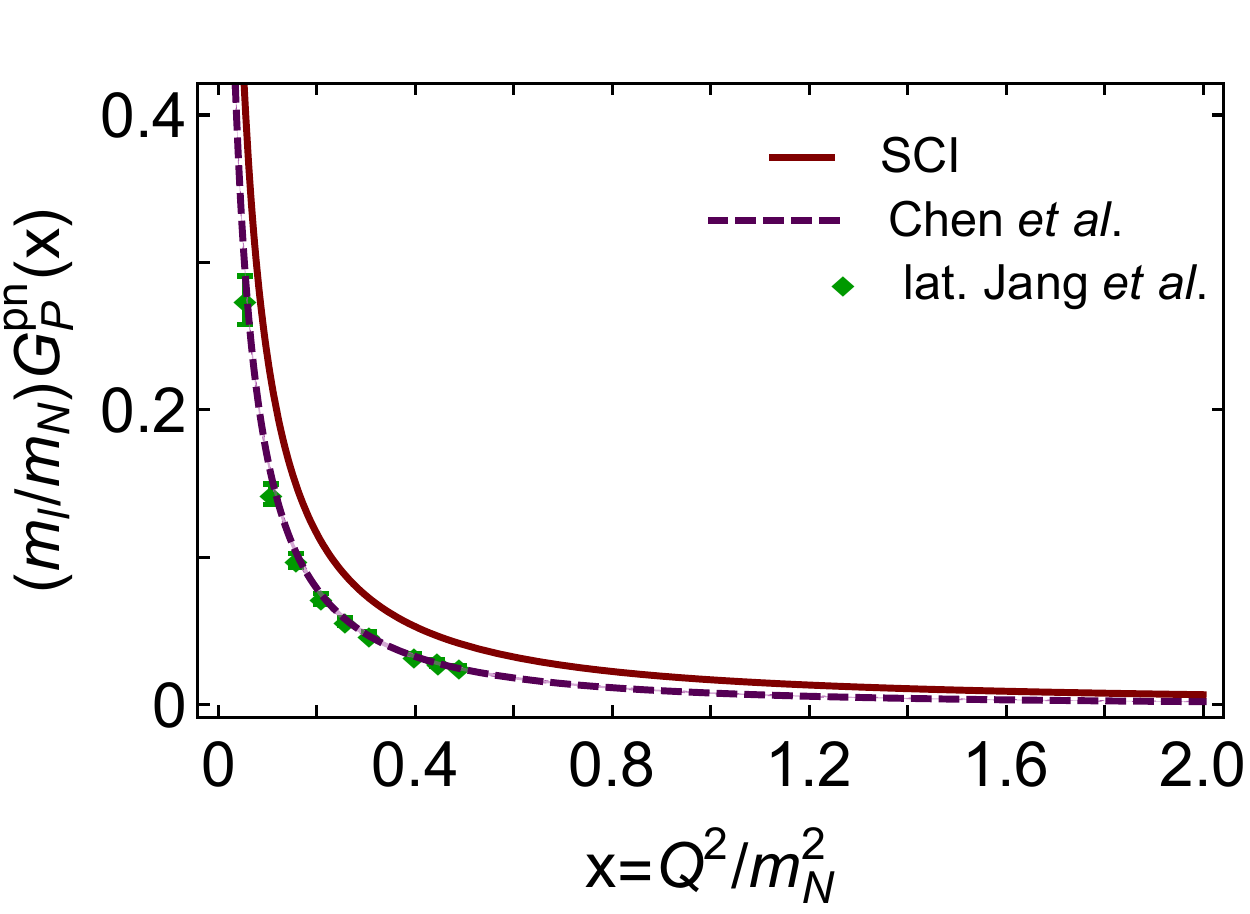}
\vspace*{1ex}

\leftline{\hspace*{0.5em}{\large{\textsf{B}}}}
\vspace*{-4ex}
\includegraphics[width=0.425\textwidth]{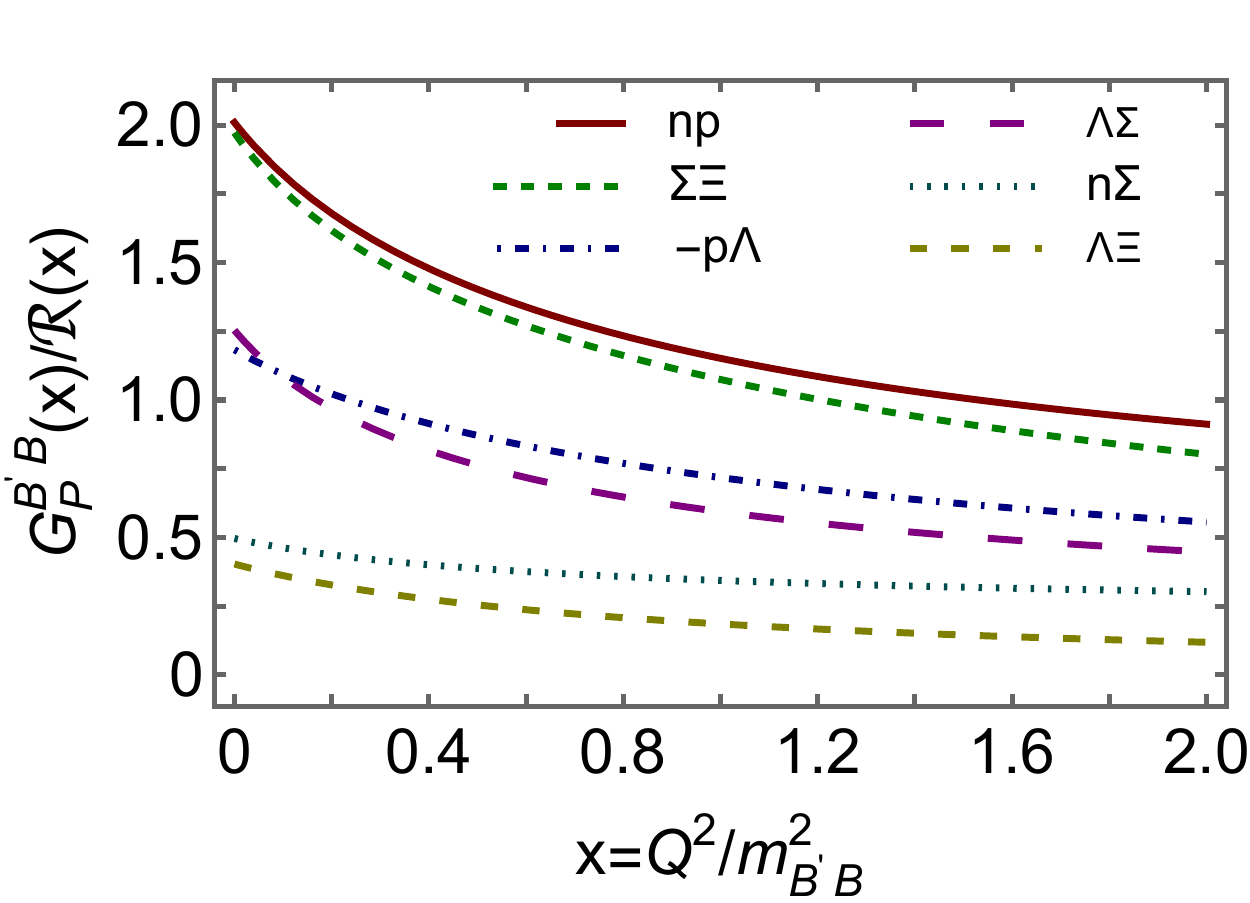}
\vspace*{1ex}

\leftline{\hspace*{0.5em}{\large{\textsf{C}}}}
\vspace*{-4ex}
\includegraphics[width=0.425\textwidth]{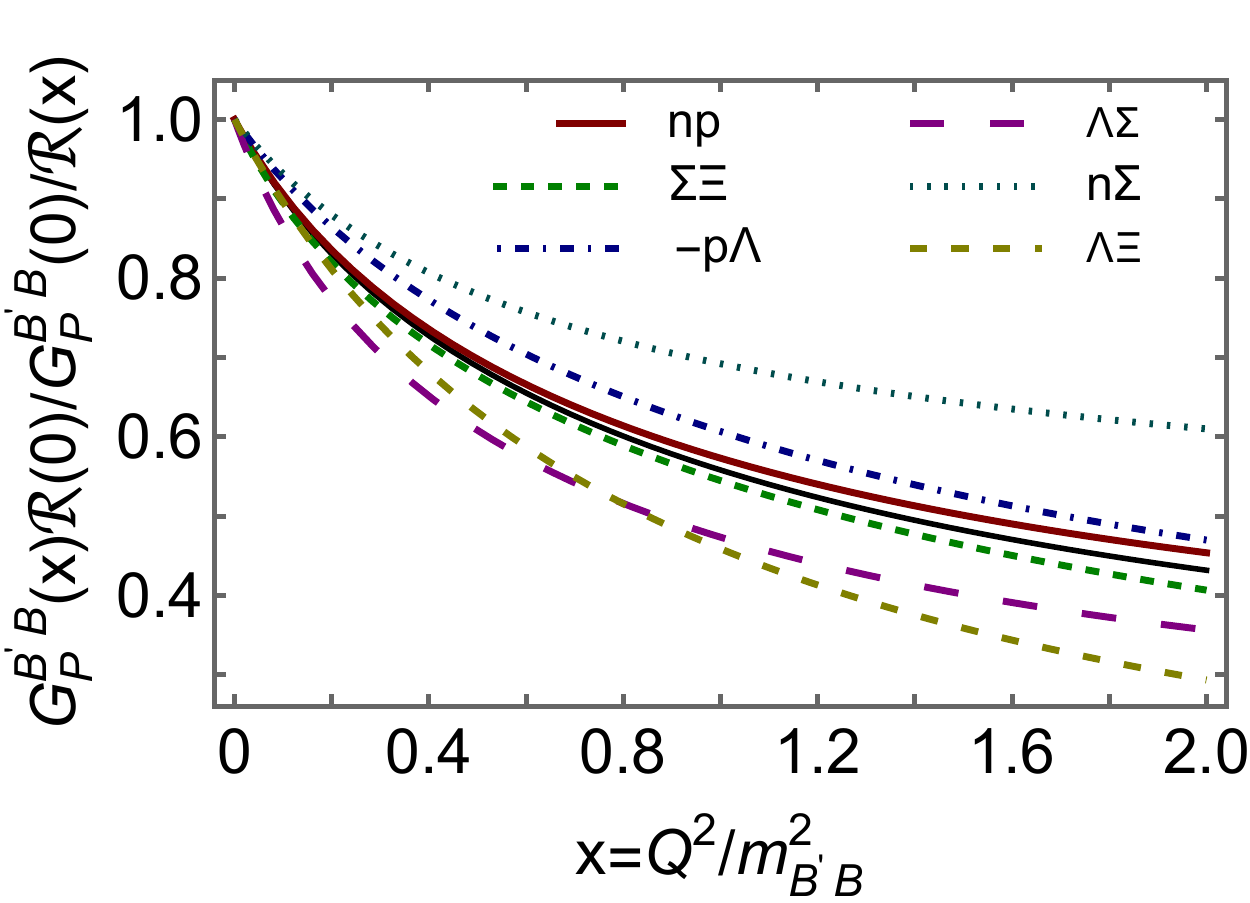}
\caption{\label{FigGPx}
{\sf Panel A}.
$(m_l/M_N)G_P^{pn}(x=Q^2/m_N^2)$: SCI result computed herein -- solid red curve; prediction from Ref.\,\cite{Chen:2021guo} -- short-dashed purple curve within like-coloured band; and lQCD results \cite{Jang:2019vkm} -- green points.
{\sf Panel B}.  Complete array of SCI predictions for octet baryon axial transition form factors: $G_P^{B^\prime B}(x=Q^2/M_{B^\prime B}^2)/{\cal R}(x)$, Eqs.\,\eqref{SCIinterpolationsGP5}, \eqref{SCIinterpolationsGP5R}.
{\sf Panel C}.  As in Panel B, but with each form factor normalised to unity at $x=0$.  The thinner solid black curve is a pointwise average of the other six curves.
}
\end{figure}

The SCI result for the $n\to p$ induced pseudoscalar transition form factor, $G_P(x)$, is reliably interpolated using the function in Eq.\,\eqref{SCIinterpolationsGP5} with the coefficients in Table~\ref{InterpolationsGA}B.  It is drawn in Fig.\,\ref{FigGPx}A and compared with both the CSM prediction from Ref.\,\cite{Chen:2021guo}, produced with QCD-like momentum dependence for all elements in Figs.\,\ref{figFaddeev}, \ref{figcurrent}, and results from a numerical simulation of lQCD \cite{Jang:2019vkm}.  Evidently, there is fair agreement between the SCI result and calculations with a closer connection to QCD.

Muon capture experiments ($\mu\,+\,p\,\to\,\nu_\mu\,+\,n$) determine the induced pseudoscalar charge
\begin{equation}
g_p^\ast = \frac{m_\mu}{2m_N}
G_p(Q^2 = 0.88\,m_\mu^2)\,,
\end{equation}
where $m_\mu$ is the muon mass.  The SCI yields $g_p^\ast=10.3$.  For comparison, we record that Ref.\,\cite{Chen:2021guo} predicts $g_p^\ast=8.80(23)$,
the MuCap Collaboration reports $g_p^\ast=8.06(55)$\,\cite{Andreev:2012fj, Andreev:2015evt},
and the world average value is $g_p^\ast=8.79(1.92)$\,\cite{Bernard:2001rs}.  Consequently, one might infer that the SCI result is $\lesssim 15$\% too large.  In assessing this outcome it is worth recalling that our SCI analysis is largely algebraic and parameter-free.

With reference to Fig.\,\ref{figcurrent}, a diagram breakdown of $G_{P}^{B^\prime B}(0)$ is presented in Table~\ref{GP0diagrams}.  One again notes the dominance of scalar diquark correlations and $0^+\leftrightarrow 1^+$ transitions in forming the induced pseudoscalar transition charges.
In these cases, however, each form factor also receives seagull contributions.
They are largest for $\Xi^- \to \Lambda$, in which the final state has all three possible types of scalar diquark correlation.  Here, the seagull terms must compensate for the strong contribution from Diagram~1.
The seagull contributions are also significant for $\Lambda\to p$ and $\Sigma^- \to n$: in the former transition they interfere constructively with Diagram~4 to compensate for a large Diagram~1 contribution; in the latter, they interfere destructively with Diagram~4.
These effects are required by PCAC and ensured by our SCI.

The full set of ground-state octet baryon induced pseudoscalar transition form factors is plotted in Fig.\,\ref{FigGPx}B.  Division by the factor ${\cal R}(x)$, defined in Eq.\,\eqref{SCIinterpolationsGP5R}, removes kinematic differences associated with quark and baryon masses and pseudoscalar meson poles.  Interpolations of these functions are provided by Eq.\,\eqref{SCIinterpolationsGP5} with the appropriate coefficients from Table~\ref{InterpolationsGA}B.
Fig.\,\ref{FigGPx}C redraws these curves renormalised to unity at $x=0$ along with the pointwise average of the rescaled functions.  On the displayed domain, the average is similar to the $n\to p$ curve; and at $x=2$, the mean absolute value of the relative deviation from the average curve is 20(14)\%.
Once again, these panels reveal that the size of SU$(3)$-flavour symmetry breaking increases with $Q^2$.  In this instance, comparing the $x=2$ values of the $n\to p$ and $\Xi^0\to \Sigma^+$ curves in Fig.\,\ref{FigGPx}C, the ratio is $\approx 1.2$; namely, alike in size with that for the axial transition form factors.

\begin{table}[t]
\caption{\label{PCouplings}
Row~1. Pseudoscalar transition couplings defined by analogy with Eq.\,\eqref{CouplingpiNN}.
Row~2. Value of this quantity at $t=0$ instead of at $t=-m_{P_{fg}}^2$.
Row~3. Relative difference between Rows~1 and 2.
}
\begin{center}
\begin{tabular*}
{\hsize}
{
l@{\extracolsep{0ptplus1fil}}
|c@{\extracolsep{0ptplus1fil}}
c@{\extracolsep{0ptplus1fil}}
c@{\extracolsep{0ptplus1fil}}
c@{\extracolsep{0ptplus1fil}}
c@{\extracolsep{0ptplus1fil}}
c@{\extracolsep{0ptplus1fil}}}\hline
& $\pi p n\ $ & $\pi \Lambda \Sigma\ $ & $K p \Lambda\ $ & $K n \Sigma\ $ & $K \Sigma \Xi\ $ & $K \Lambda \Xi\ $ \\\hline
$g_{P_{fg}B^\prime B} \frac{f_{P_{fg}}}{M_{B^\prime B}}$ &
$1.24\ $ & $0.66\ $ & $-0.83\ $ & $0.34\ $ & $1.21\ $ & $0.25\ $ \\
%
$t=0\ $ & $1.24\ $ & $0.66\ $ & $-0.82\ $ & $0.34\ $ & $1.19\ $ & $0.23\ $ \\\hline
\% difference &
$0.16\ $ & $0.15\ $ & $\phantom{-}1.5\phantom{0}\ $ & $1.8\phantom{0} $ & $1.7\phantom{0}\ $ & $9.1\phantom{0}\ $ \\
\hline
\end{tabular*}
\end{center}
\end{table}

\subsection{Pseudoscalar}
Akin to Eq.\,\eqref{CouplingKpLambda}, the $\pi N N$ form factor is defined via the pseudoscalar current in Eq.\,\eqref{jpsdq0}:
\begin{equation}
G_{\pi NN}(Q^2) \frac{f_\pi}{m_N}  \frac{m_\pi^2}{Q^2+m_\pi^2}= \frac{m_l}{m_N} G_5^{pn}(Q^2)\,.
\end{equation}
In these terms, the Goldberger-Treiman relation reads:
\begin{equation}
G_{A}^{pn}(0) = \frac{m_l}{m_N}G_5^{pn}(0)\,.
\end{equation}
Reviewing Eqs.\,\eqref{SCIinterpolations} and Table~\ref{InterpolationsGA}, it is apparent that the relation is satisfied in our SCI.
Furthermore, one can read the value of the $\pi NN$ coupling constant from the residue of $G_5^{pn}(Q^2)$ at $Q^2+m_\pi^2 = 0$:
\begin{subequations}
\begin{align}
g_{\pi NN}\frac{f_\pi}{m_N} & = \lim_{Q^2+m_\pi^2\to 0}(1+Q^2/m_\pi^2) \frac{m_l}{m_N}G_5^{pn}(Q^2) \label{CouplingpiNN}\\
& \stackrel{\rm SCI}{=} 1.24\,.
\end{align}
\end{subequations}
The SCI prediction is in fair agreement with that
obtained using QCD-kindred momentum dependence for all elements in Figs.\,\ref{figFaddeev}, \ref{figcurrent}, \emph{viz}.\ $1.29(3)$ \cite{Chen:2021guo};
extracted from pion-nucleon scattering data \cite{Baru:2011bw} -- $1.29(1)$;
inferred from the Granada 2013 $np$ and $pp$ scattering database \cite{NavarroPerez:2016eli} -- $1.30$;
and determined in a recent analysis of nucleon-nucleon scattering using effective field theory and related tools \cite{Reinert:2020mcu} -- $1.30$.

\begin{table}[t]
\caption{\label{G50diagrams}
Diagram separated contributions to $Q^2=0$ values of octet baryon pseudoscalar transition form factors, presented as a fraction of the total listed in Table~\ref{InterpolationsGA}C\,--\,Column~1 and made with reference to Fig.\,\ref{figcurrent}.
}
\begin{center}
\begin{tabular*}
{\hsize}
{
l@{\extracolsep{0ptplus1fil}}
|c@{\extracolsep{0ptplus1fil}}
c@{\extracolsep{0ptplus1fil}}
c@{\extracolsep{0ptplus1fil}}
c@{\extracolsep{0ptplus1fil}}
c@{\extracolsep{0ptplus1fil}}
c@{\extracolsep{0ptplus1fil}}}\hline
 & $\langle J \rangle^{S}_{\rm q}$  & $\langle J \rangle^{A}_{\rm q}$ &$\langle J \rangle^{AA}_{\rm qq}$ & $\langle J \rangle^{\{SA\}}_{\rm qq}$
 & $\langle J \rangle_{\rm ex}$
 & $\langle J \rangle_{\rm sg}$ \\\hline
$g_5^{pn}$ & $0.51\ $ & $\phantom{-}0.048\ $ & $\phantom{-}0.083\ $ & $0.38\ $ & $\phantom{-}0.017\ $ & $-0.039\ $ \\
$g_5^{\Sigma^- \Lambda}$ & $0.40\ $ & $\phantom{-}0.050\ $ & $\phantom{-}0.025\ $ & $0.44\ $ & $\phantom{-}0.039\ $ & $\phantom{-}0.048\ $ \\
$g_5^{p \Lambda}$ & $0.71\ $ & & $\phantom{-}0.094\ $ & $0.35\ $ & $-0.032\ $ & $-0.12\phantom{0}\ $\\
$g_5^{n\Sigma^-}$ & & $\phantom{-}0.36\phantom{0}\ $ & $-0.057\ $ & $0.59\ $ & $-0.068\ $ & $\phantom{-}0.18\phantom{0}\ $ \\
$g_5^{\Sigma^+ \Xi^0}$ & $0.57\ $ & $\phantom{-}0.028\ $ & $\phantom{-}0.073\ $ & $0.38\ $ & $-0.015\ $ & $-0.028\ $\\
$g_5^{\Lambda \Xi^-}$ & $1.49\ $ & $-0.20\phantom{0}\ $ & $\phantom{-}0.14\phantom{0}\ $ & $0.13\ $ & $\phantom{-}0.040\ $ & $-0.60\phantom{0}\ $ \\
\hline
\end{tabular*}
\end{center}
\end{table}

\begin{figure}[t]
\vspace*{2ex}

\leftline{\hspace*{0.5em}{\large{\textsf{A}}}}
\vspace*{-4ex}
\includegraphics[width=0.42\textwidth]{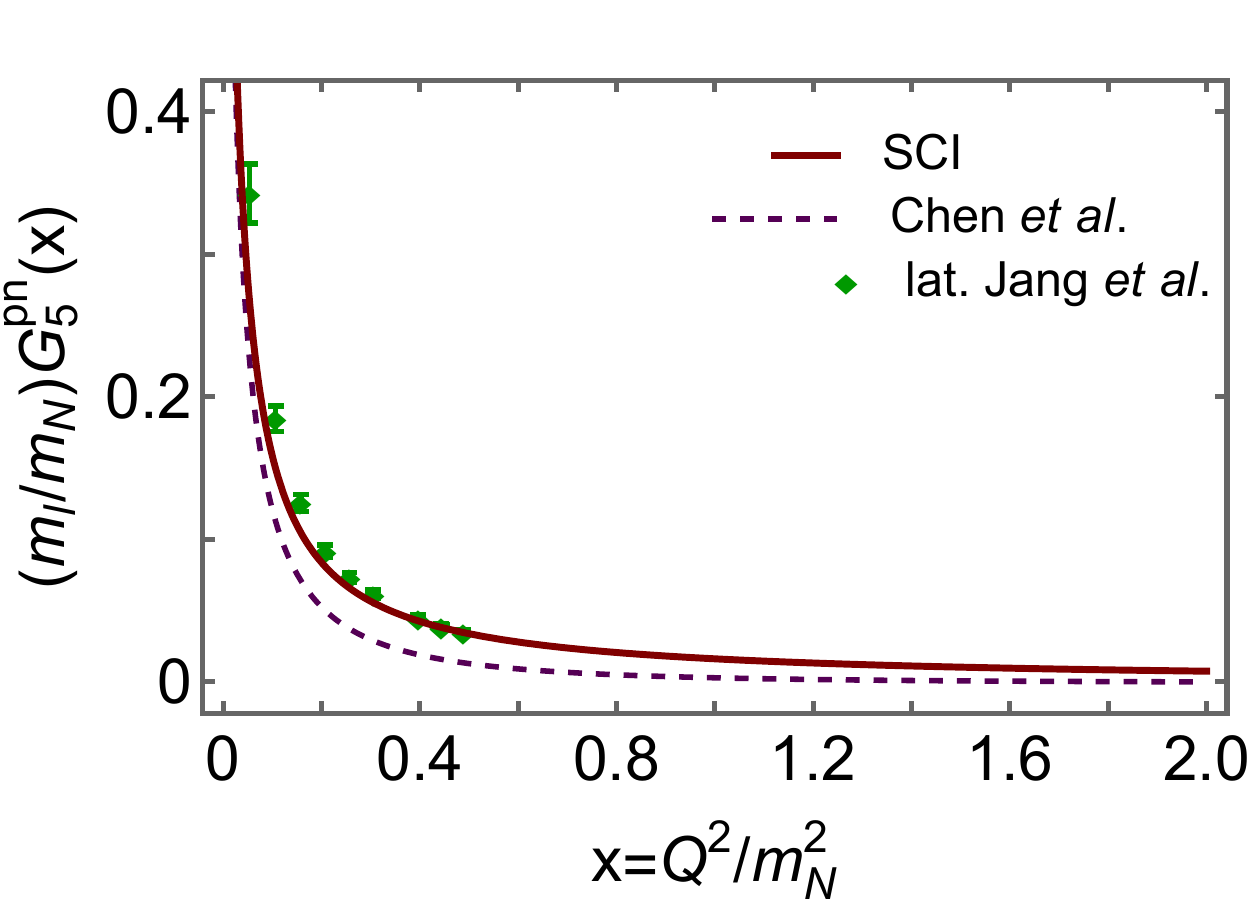}
\vspace*{1ex}

\leftline{\hspace*{0.5em}{\large{\textsf{B}}}}
\vspace*{-4ex}
\includegraphics[width=0.425\textwidth]{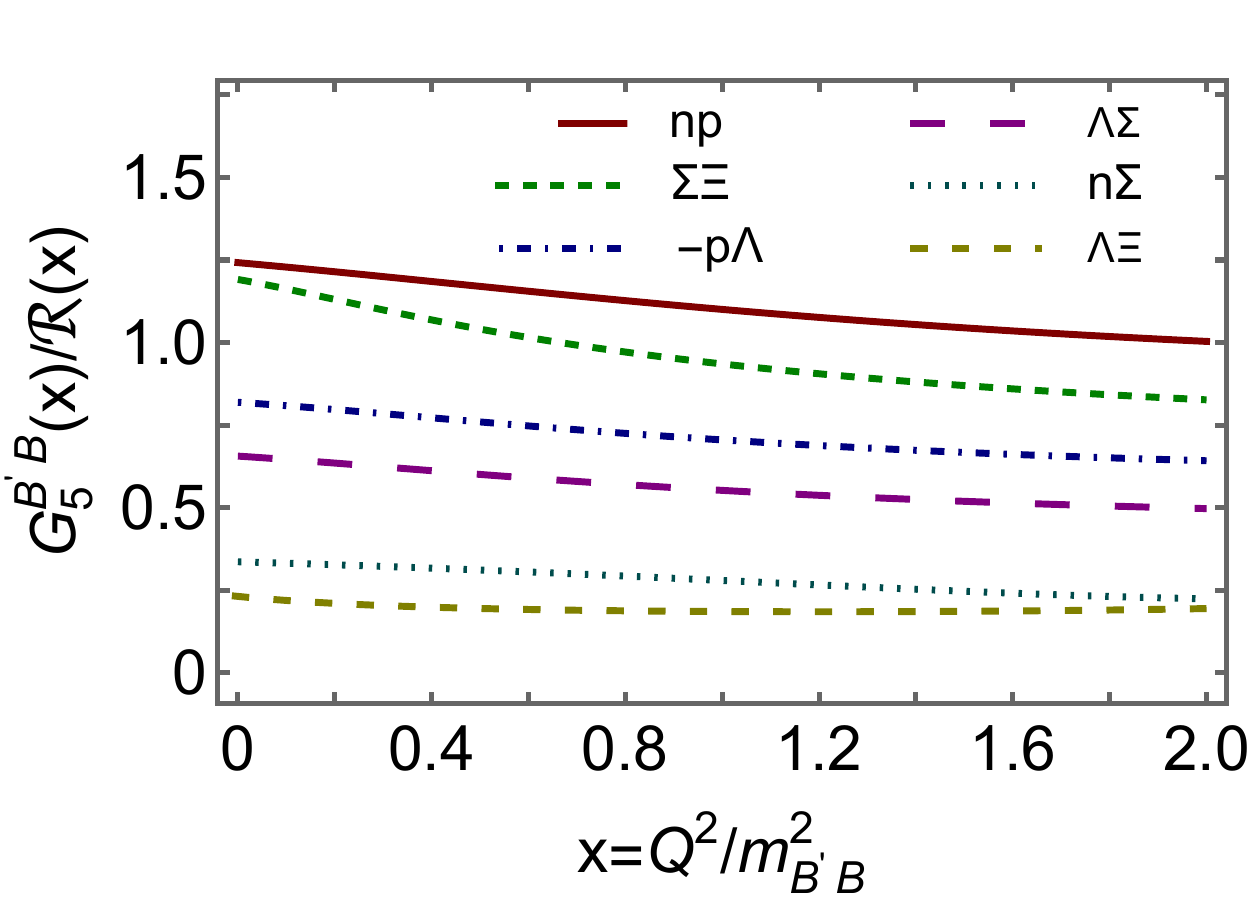}
\vspace*{1ex}

\leftline{\hspace*{0.5em}{\large{\textsf{C}}}}
\vspace*{-4ex}
\includegraphics[width=0.425\textwidth]{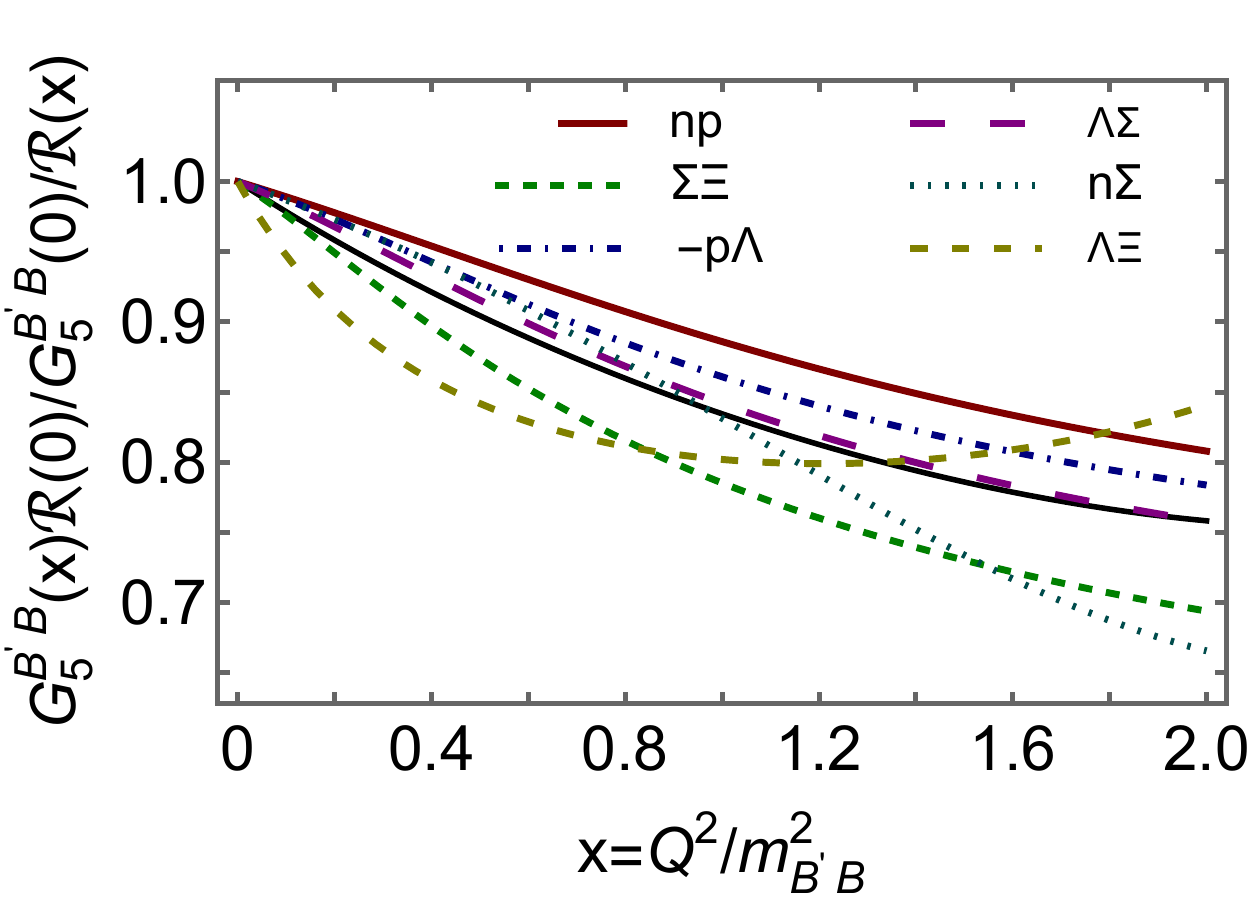}
\caption{\label{FigG5x}
{\sf Panel A}.
$(m_l/M_N)G_5^{pn}(x=Q^2/m_N^2)$: SCI result -- solid red curve; prediction from Ref.\,\cite{Chen:2021guo} -- short-dashed purple curve within like-coloured band; and lQCD results \cite{Jang:2019vkm} -- green points.
{\sf Panel B}.  Complete array of SCI predictions for octet baryon axial transition form factors: $G_P^{B^\prime B}(x=Q^2/M_{B^\prime B}^2)/{\cal R}(x)$, Eqs.\,\eqref{SCIinterpolationsGP5}, \eqref{SCIinterpolationsGP5R}.
{\sf Panel C}.  As in Panel B, but with each form factor normalised to unity at $x=0$.  The thinner solid black curve is a pointwise average of the other six curves.
}
\end{figure}

Couplings for all pseudoscalar transitions, defined by analogy with Eq.\,\eqref{CouplingpiNN}, are listed in Table~\ref{PCouplings}.  Plainly, $G_{P_{fg}B^\prime B}(0) (f_{P_{fg}}/M_{B^\prime B})$ provides a good approximation to the on-shell value of the coupling in all cases except the $\Xi^- \Lambda$ transition, which is somewhat special owing to the spin-flavour structure of the $\Lambda$, Eq.\,\eqref{SpinFlavourLambda}.  This was highlighted in connection with Table~\ref{GP0diagrams}.  Nevertheless, even in this case, the $t=0$ value is a reasonable guide.

The values in Table~\ref{PCouplings} may be compared with quark-soliton model results \cite[Table~3]{Yang:2018idi}.  Converted using empirical baryon masses and meson decay constants, the mean value of $\delta_r^g:=\{|g_{P_{fg} B^\prime B}^{\rm SCI}/g_{P_{fg} B^\prime B}^{\mbox{\rm\footnotesize \cite{Yang:2018idi}}}-1|\}$ is
$0.18(17)$.

Similar comparisons can be made with the couplings used in phenomenological hyperon+nucleon potentials \cite{Haidenbauer:2005zh, Rijken:2010zzb}, yielding $\delta_r^g=0.21(17)$, $0.15(14)$, respectively.
The dynamical coupled channels study of nucleon resonances in Ref.\,\cite{Kamano:2013iva} uses SU$(3)$-flavour symmetry to express hyperon+nucleon couplings in terms of $g_{\pi NN}$.  Relative to those couplings, one finds $\delta_r^g=0.17(15)$.  Rescaling the value of $g_{\pi NN}$ employed therein to match the SCI prediction, then $\delta_r^g=0.16(14)$.  In this last case, the difference from zero is an indication of the size of SU$(3)$-flavour symmetry violation in $\{g_{P_{fg}B^\prime B}\}$.
These comparisons with phenomenological potentials suggest that the SCI predictions for the couplings in Table~\ref{PCouplings} could serve as useful constraints in refining such models.

With reference to Fig.\,\ref{figcurrent}, a diagram breakdown of $G_{5}^{B^\prime B}(0)$ is listed in Table~\ref{G50diagrams}.  Once more, it will be observed that scalar diquark correlations dominate and $0^+\leftrightarrow 1^+$ transitions are important in building the pseudoscalar transition charges. Moreover, the pattern of diagram contributions is similar to that seen in $G_{P}^{B^\prime B}(0)$, again largely as a consequence of Eq.\,\eqref{PCAC}: recall, seagulls play no role in $G_{A}^{B^\prime B}(0)$.

The SCI result for the $n \to p$ pseudoscalar transition form factor, $G_5(x)$, is reliably interpolated using the function in Eq.\,\eqref{SCIinterpolationsGP5} with the coefficients in Table~\ref{InterpolationsGA}C.  It is plotted in Fig.\,\ref{FigG5x}A and compared with both the CSM prediction from Ref.\,\cite{Chen:2021guo}, obtained using QCD-like momentum dependence for all elements in Figs.\,\ref{figFaddeev}, \ref{figcurrent}, and results from a numerical simulation of lQCD \cite{Jang:2019vkm}.  The SCI result is harder than the CSM prediction in Ref.\,\cite{Chen:2021guo}, which should be closer to reality; hence, one may consider the possibility that the lQCD result is also too hard.

Figure~\ref{FigG5x}B depicts the complete set of ground-state octet baryon pseudoscalar transition form factors, each divided by the factor ${\cal R}(x)$ so as to remove kinematic differences associated with masses and pseudoscalar meson poles.  Interpolations of these functions are indicated by Eq.\,\eqref{SCIinterpolationsGP5} with the appropriate coefficients from Table~\ref{InterpolationsGA}C.
In Fig.\,\ref{FigG5x}C, we depict each of the curves in Panel~B after renormalisation to unity at $x=0$ alongside the pointwise average of the renormalised functions.  At $x=2$, the mean absolute value of the relative deviation from the average curve is 7(5)\%.
Focusing on the $\Xi^0\to \Sigma^+$ curves in Fig.\,\ref{FigG5x}C: at $x=2$, the ratio is $\approx 1.2$, similar to that found with $G^{\Sigma \Xi}_{A,P}$.
We note that although $G^{\Lambda\Xi^-}_5(x)/{\cal R}(x)$ is not monotonically decreasing with increasing $x$ on the domain displayed, $G^{\Lambda\Xi^-}_5(x)$ is.

\section{Valence spin fraction}
\label{SecFour}
The axial transition form factors considered above involve three distinct isospin multiplets and a singlet, which in the isospin symmetry limit may be characterised by the following four baryons: $p$, $\Sigma^+$, $\Xi^-$, $\Lambda$.  Following Ref.\,\cite{ChenChen:2022qpy}, we consider neutral-current processes and perform a flavour separation of $G_A^{B}$ in each case.  The $Q^2=0$ values of the results obtained thereby define a flavour separation of octet baryon axial charges:
{\allowdisplaybreaks
\begin{subequations}
\label{EqFlavour}
\begin{align}
g_{A}^{p} & = \phantom{-} g_{Au}^{p} - g_{Ad}^{p}\,, \\
g_{A}^{\Sigma^+} & = \phantom{-} g_{Au}^{\Sigma^+} - g_{As}^{\Sigma^+}\,, \\
g_{A}^{\Xi^-} & = -g_{Ad}^{\Xi^-} - g_{As}^{\Xi^-}\,, \\
g_{A}^{\Lambda} & =\phantom{-} g_{Au}^\Lambda + g_{Ad}^{\Lambda} - g_{As}^{\Lambda}\,.
\end{align}
\end{subequations}}

The flavour separated charges are of particular interest because $g_{Ah}^{B}$ measures the valence-$h$-quark's contribution to the light-front helicity of baryon $B$, \emph{i.e}., the difference between the light-front number-density of $h$-quarks with helicity parallel to that of the baryon and the kindred density with helicity antiparallel.
%
%
For each baryon, one may subsequently define the singlet, triplet, and octet axial charges, respectively:
\begin{subequations}
\label{axialcharges}
\begin{align}
a_0^B & = g_{Au}^{B} + g_{Ad}^{B} + g_{As}^{B} \,, \label{chargea0}\\
a_3^B & = g_{Au}^{B} - g_{Ad}^{B}  \,,\label{chargea3}\\
a_8^B & = g_{Au}^{B} + g_{Ad}^{B} - 2 g_{As}^{B} \label{chargea8} \,.
\end{align}
\end{subequations}
$a_0^B$ is the fraction of the spin of baryon $B$ that is carried by valence quarks \cite{Deur:2018roz}.  Computed in the SCI, this quantity is associated with the hadron scale, $\zeta_{\cal H} = 0.33\,$GeV \cite{Cui:2020dlm, Cui:2020tdf, Cui:2021mom, Cui:2022bxn}, whereat all properties of the hadron are carried by valence degrees of freedom.  Consequently, any difference between the SCI value of $a_0^B$ and unity should measure the fraction of the baryon's spin stored in quark+diquark orbital angular momentum.

The information provided in Appendix~\ref{AppendixSupplement} is sufficient to complete the calculation of the charges in Eq.\,\eqref{EqFlavour}.  Table~\ref{FlavourDiagramSeparated} reports the contributions to each charge from the diagrams in Fig.\,\ref{figcurrent}.  Qualitatively, the results are readily understood using the legend in Table~\ref{DiagramLegend}, the spin-flavour structure of each baryon, specified in Eqs.\,\eqref{halfnucleon}, and the Faddeev amplitudes in Table~\ref{SolveFaddeev}.
For instance, regarding the $s$-quark in the $\Lambda$:
the $s[ud]$ quark+diquark combination is strong in the Faddeev amplitude, so $g_{As}^{\Lambda}$ receives a dominant Diagram~1 scalar diquark bystander contribution;
the valence $s$ quark is never isolated alongside an axialvector diquark, hence $\langle J \rangle^{A}_{\rm s}\equiv 0$;
and the other leading contribution is from Diagram~3, which is fed by the strong $u[ds]-d[us]$ combination transforming into $u\{ds\}-d\{us\}$.
Concerning $u$ and $d$ quarks in the $\Lambda$:
the $u\leftrightarrow d$ antisymmetry of the amplitude's spin-flavour structure entails that whatever contribution $g_{Au}^\Lambda$ receives, $-g_{Ad}^\Lambda$ will be of the same size with opposite sign (weak charges of the $u$ and $d$ quarks are equal and opposite);
and diagrams involving scalar diquarks must dominate because such diquarks are most prominent in the Faddeev amplitude.

Notwithstanding the dominance of scalar diquark contributions in all cases, a material role for axialvector diquarks is also apparent.  We highlighted this with the importance of $u[ds]-d[us]\leftrightarrow u\{ds\}-d\{us\}$ in the $\Lambda$; and it is also worth emphasising the size of the $\langle J_{\rm q}^A\rangle$ contribution, which for singly-represented valence-quarks in $p$, $\Sigma^+$ is much larger in magnitude and has the opposite sign to that connected with the doubly-represented quark.

\begin{table}[t]
\caption{\label{FlavourDiagramSeparated}
With reference to Fig.\,\ref{figcurrent}, diagram contributions to flavour separated octet baryon axial charges, Eq.\,\eqref{EqFlavour}.  ``$0$'' entries are omitted.
Naturally, in the isospin-symmetry limit, the results for $\Sigma^-$ are obtained by making the replacement $g_{Au}^{\Sigma^+} \to g_{Ad}^{\Sigma^-}$; and for the $\Xi^0$, via $g_{Ad}^{\Xi^-} \to g_{Au}^{\Xi^0}$
}
\begin{center}
\begin{tabular*}
{\hsize}
{
l@{\extracolsep{0ptplus1fil}}
|c@{\extracolsep{0ptplus1fil}}
c@{\extracolsep{0ptplus1fil}}
c@{\extracolsep{0ptplus1fil}}
c@{\extracolsep{0ptplus1fil}}
c@{\extracolsep{0ptplus1fil}}
c@{\extracolsep{0ptplus1fil}}
c@{\extracolsep{0ptplus1fil}}}\hline
 & $\langle J \rangle^{S}_{\rm q}$  & $\langle J \rangle^{A}_{\rm q}$ &$\langle J \rangle^{AA}_{\rm qq}$ & $\langle J \rangle^{\{SA\}}_{\rm qq}$
 & $\langle J \rangle_{\rm ex}^{SS}$
 & $\langle J \rangle_{\rm ex}^{\{SA\}}$
 & $\langle J \rangle_{\rm ex}^{AA}$  \\\hline
$\phantom{-}g_{Au}^{p}$ & $\phantom{-}0.36\ $ & $-0.016\ $ & $\phantom{-}0.11\phantom{0}\ $ & $\phantom{-}0.22\ $ & & $\phantom{-}0.13\phantom{0}\ $& $\phantom{-}0.028\ $ \\
$-g_{Ad}^{p}$ & & $\phantom{-}0.031\ $ & $-0.022\ $ & $\phantom{-}0.22\ $ & $0.24\phantom{0}\ $ & $-0.064\ $& $\phantom{-}0.007\ $\\\hline
$\phantom{-}g_{Au}^{\Sigma^+}$ & $\phantom{-}0.40\ $ & $-0.008\ $ & $\phantom{-}0.15\phantom{0}\ $ & $\phantom{-}0.14\ $ & & $\phantom{-}0.15\phantom{0}\ $& $\phantom{-}0.023\ $ \\
$-g_{As}^{\Sigma^+}$ & & $\phantom{-}0.064\ $ & $-0.014\ $ & $\phantom{-}0.17\ $ & $0.085\ $ & $-0.014\ $& $\phantom{-}0.001\ $\\\hline
$-g_{Ad}^{\Xi^-}$ & & $\phantom{-}0.013\ $ & $-0.020\ $ & $\phantom{-}0.20\ $ & $0.24\ $ & $-0.044\ $& $\phantom{-}0.005\ $ \\
$-g_{As}^{\Xi^-}$ & $-0.61$ & $\phantom{-}0.019\ $ & $-0.066\ $ & $-0.24\ $ & & $-0.026\ $& $-0.005\ $\\\hline
$\phantom{-}g_{Au}^{\Lambda}$ & $\phantom{-}0.086\ $ & $-0.014\ $ & $\phantom{-}0.019\ $ & $-0.087\ $ & $-0.17\phantom{0}\ $ & $\phantom{-}0.035\ $& \\
$-g_{Ad}^{\Lambda}$ & $-0.086\ $ & $\phantom{-}0.014\ $ & $-0.019\ $ & $\phantom{-}0.087\ $ & $\phantom{-}0.17\phantom{0}\ $ & $-0.035\ $& \\
$-g_{As}^{\Lambda}$ & $-0.36\phantom{0}\ $ & & $-0.044\ $ & $-0.21\phantom{0}\ $ & $-0.038\ $ & $-0.016\ $& $-0.003\ $\\
\hline
\end{tabular*}
\end{center}
\end{table}

The summed results for each $g_{Af}^B$ and the associated singlet, triplet, and octet axial charges are listed in Table~\ref{AllCharges}: the pattern of the SCI predictions is similar to that in a range of other studies \cite[Table~III]{Qi:2022sus}.  Using this information, we first report the following axial charge ratios for each baryon:
\begin{equation}
\begin{array}{cccc}
g_{Ad}^p/g_{Au}^p & g_{As}^{\Sigma^+}/g_{Au}^{\Sigma^+} & g_{Ad}^{\Xi^-}/g_{As}^{\Xi^-} & g_{A(u+d)}^{\Lambda}/g_{As}^{\Lambda}\\
-0.50 & -0.34 &  -0.43 & -0.40
\end{array}\,.
\end{equation}
Evidently, the ratio of axial charges for singly and doubly represented valence quarks is roughly the same in each baryon, \emph{viz}.\ $-0.42(7)$, if one interprets $u+d$ as effectively the singly represented quark in the $\Lambda$.  Further, the ratio is smallest in magnitude when the singly represented quark is heavier than that which is doubly represented.

It is also worth recalling that the SCI produces results that are consistent with only small violations of SU$(3)$-flavour symmetry, Eq.\,\eqref{DFvalues}.  Thus, one may compare the proton results in Table~\ref{AllCharges} with the following flavour-symmetry predictions:
\begin{equation}
\frac{g_{Ad}^p}{g_{Au}^p}=\frac{F-D}{2F} = -0.39\,, \;
a_8^p=\frac{3 F-D}{F+D}=0.43\,.
\end{equation}
There is a reasonable degree of consistency.

Such accord is important because textbook-level analyses yield $g_{Ad}^p/g_{Au}^p = -1/4$ in nonrelativistic quark models with uncorrelated wave functions.
The enhanced magnitude of the SCI result can be traced to the presence of axialvector diquarks in the proton.  Namely, the fact that the Fig.\,\ref{figcurrent}\,--\,Diagram~1 contribution arising from the $\{uu\}$ correlation, in which the probe strikes the valence $d$ quark, is twice as strong as that from the $\{ud\}$, in which it strikes the valence $u$ quark.  The relative negative sign means this increases $|g_A^d|$ at a cost to $g_A^u$.  Consequently, the highly-correlated proton wave function obtained as a solution of the Faddeev equation in Fig.\,\ref{figFaddeev} places a significantly larger fraction of the proton's light-front helicity with the valence $d$ quark.

\begin{table}[t]
\caption{\label{AllCharges}
Net flavour separated and SU$(3)$ baryon axial charges obtained by combining the entries in Table~\ref{FlavourDiagramSeparated} according to Eqs.\,\eqref{EqFlavour}, \eqref{axialcharges}.
``$0$'' entries are omitted.
Recall that these results are for the elastic/neutral processes; hence, the $a_3^B$ entries need not exactly match those in Row~1 of Table~\ref{GAzero}.
}
\begin{center}
\begin{tabular*}
{\hsize}
{
l@{\extracolsep{0ptplus1fil}}
|c@{\extracolsep{0ptplus1fil}}
c@{\extracolsep{0ptplus1fil}}
c@{\extracolsep{0ptplus1fil}}
c@{\extracolsep{0ptplus1fil}}
c@{\extracolsep{0ptplus1fil}}
c@{\extracolsep{0ptplus1fil}}
c@{\extracolsep{0ptplus1fil}}}\hline
$B \ $ & $g_{Au}^B\ $ & $g_{Ad}^B\ $ & $g_{As}^B\ $ & $a_{0}^B\ $ & $a_{3}^B\ $ & $a_{8}^B\ $ \\\hline
$p\ $ & $\phantom{-}0.83\ $ & $-0.41\ $ & & $0.42\ $ & $1.24\ $ & $\phantom{-}0.42\ $ \\
$\Sigma^+\ $ & $\phantom{-}0.85\ $ & & $-0.29\ $ & $0.56\ $ & $0.85\ $ & $\phantom{-}1.42\ $ \\
$\Xi^-\ $ & & $-0.40\ $ & $\phantom{-}0.93\ $ & $0.53\ $ & $0.40\ $ & $-2.26\ $ \\
$\Lambda\ $ & $-0.13\ $ & $-0.13\ $ & $\phantom{-}0.67\ $ & $0.41\ $ & & $-1.61\ $ \\
\hline
\end{tabular*}
\end{center}
\end{table}

The enhancement remains when all elements in Figs.\,\ref{figFaddeev}, \ref{figcurrent} express QCD-like momentum dependence, but with reduced magnitude \cite{Chen:2021guo}: $g_{Ad}^p/g_{Au}^u = - 0.32(2)$.  Relative to that analysis, the larger size of the SCI result likely owes to the momentum-independence of the Bethe-Salpeter and Faddeev amplitudes it generates.  This limits the suppression of would-be soft contributions, \emph{e.g}., the two-loop $\langle J \rangle^{SS}_{\rm ex}$ contribution in Row~2 of Table~\ref{FlavourDiagramSeparated} is roughly five-times larger than the analogous term in Ref.\,\cite{Chen:2021guo}, significantly enhancing the magnitude of $g_{Ad}^p$.

Referring to Table~\ref{AllCharges}, $a_3^B$ and $a_8^B$ are conserved charges, \emph{i.e}., they are the same at all resolving scales, $\zeta$.  This is not true of the individual terms in their definitions, Eqs.\,\eqref{chargea3}, \eqref{chargea8}: the flavour-separated valence quark charges $g_{Au}^B$, $g_{Ad}^B$, $g_{As}^B$ evolve with $\zeta$ \cite{Deur:2018roz}.  Consequently, the value of $a_0^B$, which is identified with the fraction of the baryon's total $J=1/2$ carried by its valence degrees-of-freedom, changes with scale -- it diminishes slowly with increasing $\zeta$; and as noted above, the SCI predictions in Table~\ref{AllCharges} are made with reference to the hadron scale $\zeta=\zeta_{\cal H} = 0.33\,$GeV \cite{Cui:2020dlm, Cui:2020tdf, Cui:2021mom, Cui:2022bxn}.

Textbook-level analyses yield $a_0^B=1$ in nonrelativistic quark models with uncorrelated wave functions.  So, in such pictures, all the baryon's spin derives from that of the constituent quarks.  Herein, on the other hand, considering the hadron scale, then the valence degrees-of-freedom in octet baryons carry roughly one-half the total spin.  The mean is
\begin{equation}
\bar a_0^B = 0.50(7)\,.
\end{equation}
Since there are no other degrees-of-freedom at this scale and the Poincar\'e-covariant baryon wave function obtained from the Faddeev amplitude discussed in Appendix~\ref{AppendixFaddeev} properly describes a $J=1/2$ system, then the remainder of the total-$J$ must be lodged with quark+diquark orbital angular momentum.  In keeping with such a picture, this remainder is largest in systems with the lightest valence degrees-of-freedom: $a_0^p \approx a_0^\Lambda < a_0^{\Sigma} \approx a_0^{\Xi}$.  A detailed discussion of these and related issues will be presented elsewhere \cite{Peng:2022Progress}.

\section{Summary and Perspective}
\label{epilogue}
Using a symmetry-preserving treatment of a vector$\,\times\,$vector contact interaction (SCI), we delivered predictions for the axial, induced-pseudoscalar, and pseudoscalar transition form factors of ground state octet baryons, thereby furthering progress toward a goal of unifying an array of baryon properties \cite{Wilson:2011aa, Xu:2015kta, Yin:2021uom, Raya:2021pyr} with analogous treatments of semileptonic decays of heavy+heavy and heavy+light pseudoscalar mesons to both pseudoscalar and vector meson final states \cite{Xu:2021iwv, Xing:2022sor}.
The study required an extensive body of calculations, demanding solutions of a collection of integral equations for an array of relevant $n=2$-$6$--point Schwinger functions, \emph{e.g}., gap, Bethe-Salpeter, and, of special importance, Faddeev equations that describe octet baryons as quark--plus--interacting-diquark bound-states.
Naturally, being symmetry-preserving, all mathematical and physical expressions of partial conservation of the axial current (PCAC) are manifest.

Our implementation of the SCI has four parameters, \emph{viz}.\ the values of a mass-dependent quark+antiquark coupling strength chosen at the current-masses of the $u/d$, $s$, $c$, $b$ quarks.  Since their values were fixed elsewhere \cite{Xu:2021iwv}, the predictions for octet baryons presented herein are parameter free.
The merits of the SCI are its
algebraic simplicity;
paucity of parameters;
simultaneous applicability to a wide variety of systems and processes;
and potential for revealing insights that connect and explain numerous phenomena.

Regarding octet baryon axial transition form factors, $G_A$, SCI results are consistent with a small violation of SU$(3)$-flavour symmetry [Sec.\,\ref{GAsection}]; and our analysis revealed this outcome to be a dynamical consequence of emergent hadron mass.  Namely, the generation of a nuclear size mass-scale in the strong interaction sector of the Standard Model acts to mask the impact of Higgs-boson generated differences between the current masses of lighter quarks.
Furthermore, the spin-flavour structure of the Poincar\'e-covariant baryon wave functions, expressed in the presence of both flavour-antitriplet scalar diquarks and flavour-sextet axialvector diquarks, plays a key role in determining the axial charges and form factors.  Notably, whilst scalar diquark contributions are dominant, axial vector diquarks nevertheless play a material role, which is especially visible in the values of the flavour-separated charges.  Thus here, as with many other quantities \cite{Chang:2022jri, Lu:2022cjx, Cui:2021gzg}, a sound description of observables requires the presence of axialvector correlations in the wave functions of ground-state octet baryons.

Octet baryon induced-pseudoscalar transition form factors, $G_P$, are also described well by our SCI [Sec.\,\ref{SecGP}].  Qualitatively, the same formative elements are at work with $G_P$ as with $G_A$.  The material difference is the role of seagull terms in the current [Fig.\,\ref{figcurrent}].  $G_A$ is associated with the transverse part of the baryon axial current; hence, receives no seagull contributions.  On the other hand, seagull terms contribute to all calculated induced pseudoscalar form factors, being particularly significant for $\Xi^- \to \Lambda$, $\Lambda\to p$ and $\Sigma^- \to n$.  Each $G_P(Q^2)$ exhibits a pole at $Q^2+m_{\cal P}^2$, where $m_{\cal P}= m_\pi, m_K$, the pion or kaon mass, depending on whether the underlying weak quark transition is $d\to u$ or $s\to u$.

Owing to PCAC, which entails that the longitudinal part of the axialvector current is completely determined by the kindred pseudoscalar form factor, then in every case there is an intimate connection between the induced pseudoscalar and pseudoscalar transition form factors, $G_{P,5}$.  Consequently, viewed from the correct perspective, all said about $G_P$ applies equally to $G_5$.  A new feature is the link between $G_5$ and a number of meson+baryon couplings, which can be read from the residue of $G_5$ at $Q^2+m_{\cal P}^2=0$ [Table~\ref{PCouplings}].  Thus computed, the SCI prediction for the $\pi pn $ coupling is in fair agreement with other calculations and phenomenology.

Working with neutral axial currents, we obtained SCI predictions for the flavour separation of octet baryon axial charges and, therefrom, values for the associated SU$(3)$-flavour singlet, triplet and octet axial charges [Sec.\,\ref{SecFour}].  The singlet charge relates to the fraction of a baryon's total angular momentum carried by its valence quarks.  The SCI predicts that, at the hadron scale, $\zeta_{\cal H}=0.33\,$GeV, this fraction is roughly 50\%.  Since there are no other degrees-of-freedom at $\zeta_{\cal H}$, the remainder may be associated with quark+diquark orbital angular momentum.

Numerous analyses have shown that when viewed prudently, SCI results typically provide a useful quantitative guide.  Notwithstanding this, it is worth checking the predictions described herein using the QCD-kindred framework that has been employed widely in studying properties of the nucleon, $\Delta$-baryon, and their low-lying excitations \cite{Chen:2018nsg, Lu:2019bjs, Cui:2020rmu, Liu:2022ndb, Chen:2020wuq, Chen:2021guo, ChenChen:2022qpy}.  This is especially true of the results for octet baryon spin structure.
In addition, with continuing progress in developing the \emph{ab initio} Poincar\'e-covariant three-body Faddeev equation approach to baryon structure \cite{Eichmann:2009qa, Eichmann:2011pv, Wang:2018kto, Qin:2019hgk}, it should soon be possible to deliver octet baryon axial and pseudoscalar current form factors independently of the quark+diquark scheme.  Comparisons between the results obtained in the different frameworks should serve to improve both.
Naturally, too, an extension of the analyses herein to baryons containing one or more heavy quarks would also be valuable; especially, \emph{e.g}., given the role that $\Lambda_b \to \Lambda_c e^- \bar\nu_e$ may play in testing lepton flavour universality \cite{Li:2021qod}.

\begin{acknowledgments}
We are grateful for constructive comments from D.\,S.~Carman, R.\,W.~Gothe, G.~Krein, T.-S.\,H.~Lee, V.\,I.~Mokeev, H.-Y.~Xing, Z.-N. Xu and \mbox{F.-S.~Yu}.
Work supported by:
National Natural Science Foundation of China (grant nos.\,12135007, 12047502);
and
Natural Science Foundation of Jiangsu Province (grant no.\ BK20220122).
%
%
\end{acknowledgments}

\appendix

\section{SCI Propagators, Amplitudes, and Currents}
\label{AppendixSupplement}
\subsection{Contact Interaction}
\label{AppendixSCI}
The basic element in the continuum analysis of hadron bound states is the quark+antiquark scattering kernel.  At leading-order in a widely-used symmetry-preserving approximation scheme (rainbow-ladder -- RL -- truncation) \cite{Munczek:1994zz, Bender:1996bb}, it can be written:
\begin{subequations}
\label{KDinteraction}
\begin{align}
\mathscr{K}_{\alpha_1\alpha_1',\alpha_2\alpha_2'}  & = {\mathpzc G}_{\mu\nu}(k) [i\gamma_\mu]_{\alpha_1\alpha_1'} [i\gamma_\nu]_{\alpha_2\alpha_2'}\,,\\
 {\mathpzc G}_{\mu\nu}(k)  & = \tilde{\mathpzc G}(k^2) T^k_{\mu\nu}\,,
\end{align}
\end{subequations}
where $k = p_1-p_1^\prime = p_2^\prime -p_2$, with $p_{1,2}$, $p_{1,2}^\prime$ being, respectively, the initial and final momenta of the scatterers, and $k^2T_{\mu\nu}^k = k^2\delta_{\mu\nu} - k_\mu k_\nu$.

$\tilde{\mathpzc G}$ is the defining element; and it is now known that, owing to the emergence of a gluon mass-scale \cite{Boucaud:2011ug, Aguilar:2015bud, Gao:2017uox, Cui:2019dwv}, $\tilde{\mathpzc G}$ is nonzero and finite at infrared momenta.  Hence, it can be written as follows:
\begin{align}
\tilde{\mathpzc G}(k^2) & \stackrel{k^2 \simeq 0}{=} \frac{4\pi \alpha_{\rm IR}}{m_G^2}\,.
\end{align}
In QCD \cite{Cui:2019dwv}: $m_G \approx 0.5\,$GeV, $\alpha_{\rm IR} \approx \pi$.
Following Ref.\,\cite{Xu:2021iwv}, we retain this value of $m_G$ and, exploiting the fact that a SCI cannot support relative momentum between meson bound-state constituents, simplify the tensor in Eqs.\,\eqref{KDinteraction}:
\begin{align}
\label{KCI}
\mathscr{K}_{\alpha_1\alpha_1',\alpha_2\alpha_2'}^{\rm CI}  & = \frac{4\pi \alpha_{\rm IR}}{m_G^2}
 [i\gamma_\mu]_{\alpha_1\alpha_1'} [i\gamma_\mu]_{\alpha_2\alpha_2'}\,.
 \end{align}

An elementary form of confinement is expressed in the SCI by including an infrared regularising scale, $\Lambda_{\rm ir}$, when defining all integral equations relevant to bound-state problems \cite{Ebert:1996vx}.  This expedient excises momenta below $\Lambda_{\rm ir}$, and so eliminates quark+antiquark production thresholds \cite{Krein:1990sf}.  The standard choice is $\Lambda_{\rm ir} = 0.24\,$GeV\,$=1/[0.82\,{\rm fm}]$ \cite{GutierrezGuerrero:2010md}, which introduces a confinement length scale that is roughly the same as the proton radii \cite{Cui:2022fyr}.

All integrals in SCI bound-state equations require ultraviolet regularisation.  This step breaks the link between infrared and ultraviolet scales that is characteristic of QCD.  Consequently, the associated ultraviolet mass-scales, $\Lambda_{\rm uv}$, become physical parameters.  They may be interpreted as upper bounds on the domains whereupon distributions within the associated systems are practically momentum-independent.

{\allowdisplaybreaks
For a quark of flavour $f$, the SCI gap equation is
\begin{align}
\label{GapEqn}
S_f^{-1}(p)  & = i\gamma\cdot p +m_f \nonumber \\
& \quad + \frac{16 \pi}{3} \frac{\alpha_{\rm IR}}{m_G^2}
\int \frac{d^4q}{(2\pi)^4} \gamma_\mu S_f(q) \gamma_\mu\,,
\end{align}
where $m_f$ is the $f$-quark current-mass.  Using a Poincar\'e-invariant regularisation, the solution is
\begin{equation}
\label{genS}
S_f^{-1}(p) = i \gamma\cdot p + M_f\,,
\end{equation}
with $M_f$, the dynamically generated dressed-quark mass, obtained as the solution of
\begin{equation}
M_f = m_f + M_f\frac{4\alpha_{\rm IR}}{3\pi m_G^2}\,\,{\cal C}_0^{\rm iu}(M_f^2)\,,
\label{gapactual}
\end{equation}
where  ($\tau_{\rm uv}^2=1/\Lambda_{\textrm{uv}}^{2}$, $\tau_{\rm ir}^2=1/\Lambda_{\textrm{ir}}^{2}$)
\begin{align}
\nonumber
{\cal C}_0^{\rm iu}(\sigma) &=
\int_0^\infty\! ds \, s \int_{\tau_{\rm uv}^2}^{\tau_{\rm ir}^2} d\tau\,{\rm e}^{-\tau (s+\sigma)}\\
& =
\sigma \big[\Gamma(-1,\sigma \tau_{\rm uv}^2) - \Gamma(-1,\sigma \tau_{\rm ir}^2)\big].
\label{eq:C0}
\end{align}
The ``iu'' superscript stresses that the function depends on both the infrared and ultraviolet cutoffs and
$\Gamma(\alpha,y)$ is the incomplete gamma-function.
In general, functions of the following type arise in SCI bound-state equations:
\begin{align}
%
%
%
\overline{\cal C}^{\rm iu}_n(\sigma) & = \Gamma(n-1,\sigma \tau_{\textrm{uv}}^{2}) - \Gamma(n-1,\sigma \tau_{\textrm{ir}}^{2})\,,
\label{eq:Cn}
\end{align}
${\cal C}^{\rm iu}_n(\sigma)=\sigma \overline{\cal C}^{\rm iu}_n(\sigma)$, $n\in {\mathbb Z}^\geq$.}

The SCI analysis of pseudoscalar mesons in Ref.\,\cite{Xu:2021iwv} improved upon that in Ref.\,\cite{Roberts:2011wy} by keeping all light-quark parameter values therein but fixing the $s$-quark current mass, $m_s$, and $K$-meson ultraviolet cutoff, $\Lambda_{\rm uv}^K$, through a least-squares fit to measured values of $m_K$, $f_K$, whilst imposing the relation:
\begin{equation}
\alpha_{\rm IR}(\Lambda_{\rm uv}^{K}) [\Lambda_{\rm uv}^{K}]^2 \ln\frac{\Lambda_{\rm uv}^{K}}{\Lambda_{\rm ir}}
=
\alpha_{\rm IR}(\Lambda_{\rm uv}^{\pi}) [\Lambda_{\rm uv}^{\pi}]^2 \ln\frac{\Lambda_{\rm uv}^{\pi}}{\Lambda_{\rm ir}}\,.
\label{alphaLambda}
\end{equation}
This procedure eliminates one parameter by imposing the physical constraint that any increase in the momentum-space extent of a hadron wave function should be matched by a reduction in the effective coupling between the constituents.  We use the $u/d$, $s$ values herein.  The procedure was also implemented for the $c$-quark/$D$-meson and $\bar b$-quark/$B$-meson; and the complete set of results is reproduced in Table~\ref{Tab:DressedQuarks}.
The evolution of $\Lambda_{\rm uv}$ with $m_P$ is described by the following interpolation $(s=m_{P}^2)$:
\begin{equation}
\label{LambdaIRMass}
\Lambda_{\rm uv}(s) = 0.306 \ln [ 19.2 + (s/m_\pi^2-1)/2.70]\,.
\end{equation}

\begin{table}[t]
\caption{\label{Tab:DressedQuarks}
Couplings, $\alpha_{\rm IR}/\pi$, ultraviolet cutoffs, $\Lambda_{\rm uv}$, and current-quark masses, $m_f$, $f=u/d,s,c,b$, that deliver a good description of flavoured pseudoscalar meson properties, along with the dressed-quark masses, $M$, and pseudoscalar meson masses, $m_{P}$, and leptonic decay constants, $f_{P}$, they produce; all obtained with $m_G=0.5\,$GeV, $\Lambda_{\rm ir} = 0.24\,$GeV.
Empirically, at a sensible level of precision \cite{Zyla:2020zbs}:
$m_\pi =0.14$, $f_\pi=0.092$;
$m_K=0.50$, $f_K=0.11$;
$m_{D} =1.87$, $f_{D}=0.15$;
$m_{B}=5.30$, $f_{B}=0.14$.
%
(We assume isospin symmetry and list dimensioned quantities in GeV.)}
\begin{center}
\begin{tabular*}
{\hsize}
{
l@{\extracolsep{0ptplus1fil}}|
c@{\extracolsep{0ptplus1fil}}|
c@{\extracolsep{0ptplus1fil}}
c@{\extracolsep{0ptplus1fil}}
c@{\extracolsep{0ptplus1fil}}
|c@{\extracolsep{0ptplus1fil}}
c@{\extracolsep{0ptplus1fil}}
c@{\extracolsep{0ptplus1fil}}}\hline
& quark & $\alpha_{\rm IR}/\pi\ $ & $\Lambda_{\rm uv}$ & $m$ &   $M$ &  $m_{P}$ & $f_{P}$ \\\hline
$\pi\ $  & $l=u/d\ $  & $0.36\phantom{2}$ & $0.91\ $ & $0.0068_{u/d}\ $ & 0.37$\ $ & 0.14 & 0.10  \\\hline
$K\ $ & $\bar s$  & $0.33\phantom{2}$ & $0.94\ $ & $0.16_s\phantom{7777}\ $ & 0.53$\ $ & 0.50 & 0.11 \\\hline
$D\ $ & $c$  & $0.12\phantom{2}$ & $1.36\ $ & $1.39_c\phantom{7777}\ $ & 1.57$\ $ & 1.87 & 0.15 \\\hline
$B\ $ & $\bar b$  & $0.052$ & $1.92\ $ & $4.81_b\phantom{7777}\ $ & 4.81$\ $ & 5.30 & 0.14
\\\hline
\end{tabular*}
\end{center}
\end{table}

\subsection{Diquarks}
\label{Appendixdiquarks}
One now has all information necessary to specify the dressed-quark propagators that appear when solving Fig.\,\ref{figFaddeev} for octet baryons.  The next step is to compute the SCI diquark correlation amplitudes.  The forms of the relevant Bethe-Salpeter equations are written in Ref.\,\cite[Sec.\,2.2.2]{Chen:2012qr}, along with the structure of their solutions, which can be expressed as follows:
\begin{equation}
\label{DefineBSAs}
^a\Gamma_{fg}^{J^P}(K)
= T_{\bar 3_c}^a \otimes \underline\Gamma_{fg}^{J^P}(K)
= T_{\bar 3_c}^a \otimes t^J_{fg}\otimes\Gamma_{fg}^{J^P}(K)\,,
\end{equation}
where the colour-antitriplet character is expressed in $\{T_{\bar 3_c}^a,a=1,2,3\}=\{i\lambda^2, i\lambda^5, i\lambda^7\}$, using Gell-Mann matrices;
the flavour structure is expressed via
{\allowdisplaybreaks
\begin{align}
t_{ud}^0 =
\left[\begin{array}{ccc}
0 & 1 & 0 \\
-1 & 0 & 0 \\
0 & 0 & 0
\end{array}\right], &
\; t_{us}^0 =
\left[\begin{array}{ccc}
0 & 0 & 1 \\
0 & 0 & 0 \\
-1 & 0 & 0
\end{array}\right] \,, \nonumber\\
t_{ds}^0 =
\left[\begin{array}{ccc}
0 & 0 & 0 \\
0 & 0 & 1 \\
0 & -1 & 0
\end{array}\right], &
\,\nonumber\\
t_{uu}^1 =
\left[\begin{array}{ccc}
\surd 2 & 0 & 0 \\
0 & 0 & 0 \\
0 & 0 & 0
\end{array}\right], &
\; t_{ud}^1 =
\left[\begin{array}{ccc}
0 & 1 & 0 \\
1 & 0 & 0 \\
0 & 0 & 0
\end{array}\right], \nonumber \\
t_{us}^1 =
\left[\begin{array}{ccc}
0 & 0 & 1 \\
0 & 0 & 0 \\
1 & 0 & 0
\end{array}\right], &
\; t_{dd}^1 =
\left[\begin{array}{ccc}
0 & 0 & 0 \\
0 & \surd 2 & 0 \\
0 & 0 & 0
\end{array}\right], \nonumber\\
t_{ds}^1 =
\left[\begin{array}{ccc}
0 & 0 & 0 \\
0 & 0 & 1 \\
0 & 1 & 0
\end{array}\right],  &
\; t_{ss}^1 =
\left[\begin{array}{ccc}
0 & 0 & 0 \\
0 &0 & 0 \\
0 & 0 & \surd 2
\end{array}\right];
\end{align}
and the Dirac structure in
\begin{subequations}
\label{qqBSAs}
\begin{align}
\Gamma_{fg}^{0^+}(K) & =
\gamma_5\left[ i E_{[fg]} + \frac{\gamma\cdot K}{2 M_{R}} F_{[fg]} \right]C,\\
\Gamma_{fg}^{1^+}(K) & = T_{\mu\nu}^K \gamma_\nu C E_{\{fg\}}\,,
\end{align}
\end{subequations}
where $K$ is the correlation's total momentum,
$M_{R}= M_f M_g/[M_f+M_g]$,
and $C=\gamma_2\gamma_4$ is the charge conjugation matrix.
As initially observed in Ref.\,\cite{Cahill:1987qr}, owing to similarities between their respective Bethe-Salpeter equations, one may consider a colour-antitriplet $J^P$ diquark as being the partner to a colour-singlet $J^{-P}$ meson.  Thus, the $J^P$ diquark Bethe-Salpeter equations are solved using the dressed-quark propagators described above and the values of $\Lambda_{\rm uv}$ associated with the $J^{-P}$ mesons \cite{Yin:2019bxe, Yin:2021uom}.  The calculated diquark masses and canonically normalised amplitudes required herein are listed in Table~\ref{qqBSAsolutions}.
(As explained in Ref.\,\cite[Appendix~C]{Chen:2012txa}, when using the SCI it is necessary to slightly modify the canonical normalisation procedure for a given diquark correlation amplitude, resulting in a $\lesssim 4$\% recalibration, which is already included in Table~\ref{qqBSAsolutions}.)
}

\begin{table}[t]
\caption{\label{qqBSAsolutions}
Masses and canonically normalised correlation amplitudes obtained by solving the diquark Bethe-Salpeter equations.  Recall that we work in the isospin-symmetry limit.
(Masses listed in GeV.  Amplitudes are dimensionless.)}
\begin{center}
\begin{tabular*}
{\hsize}
{
c@{\extracolsep{0ptplus1fil}}
c@{\extracolsep{0ptplus1fil}}
c@{\extracolsep{0ptplus1fil}}|
c@{\extracolsep{0ptplus1fil}}
c@{\extracolsep{0ptplus1fil}}
c@{\extracolsep{0ptplus1fil}}}\hline
$m_{[ud]}$ & $E_{[ud]}$ & $F_{[ud]}\ $ & $m_{[us]}$ & $E_{[us]}$ & $F_{[us]}\ $ \\
$0.78$ & $2.71$ &  $0.31\ $ &  $0.94$  & $2.78$ & $0.37\ $
\end{tabular*}
\begin{tabular*}
{\hsize}
{
c@{\extracolsep{0ptplus1fil}}
c@{\extracolsep{0ptplus1fil}}
|c@{\extracolsep{0ptplus1fil}}
c@{\extracolsep{0ptplus1fil}}
|c@{\extracolsep{0ptplus1fil}}
c@{\extracolsep{0ptplus1fil}}}\hline
$m_{\{uu\}}$ & $E_{\{uu\}}\ $ & $m_{\{us\}}$ & $E_{\{us\}}\ $ & $m_{\{ss\}}$ & $E_{\{ss\}}\ $ \\
$1.06$ & $1.39\ $ &  $1.22$ &  $1.16\ $  & $1.33$ & $1.10\ $\\\hline
\end{tabular*}
\end{center}
\end{table}

The scalar and axialvector diquark propagators take standard forms:
\begin{subequations}
\label{qqPropagator}
\begin{align}
\Delta^{[fg]}(K) & = \frac{1}{K^2 + m_{[fg]}^2}\,, \\
\Delta^{\{fg\}}_{\mu\nu}(K) & = \left[\delta_{\mu\nu}+\frac{K_\mu K_\nu}{m_{\{fg\}}^2}\right] \frac{1}{K^2 + m_{\{fg\}}^2}\,,
\end{align}
\end{subequations}
where the masses are taken from Table~\ref{qqBSAsolutions}.

\subsection{Faddeev amplitudes}
\label{AppendixFaddeev}
All elements necessary to compose the octet baryon Faddeev kernels are now in hand and we complete this task following Ref.\,\cite[Sec.\,3]{Chen:2012qr}.  The value of $\Lambda_{\rm uv}$ in each Faddeev equation is chosen to be the scale associated with the lightest diquark in the bound-state because this is always the smallest value; hence, the dominant regularising influence.

Any $J=1/2^+$ octet solution of the resulting Faddeev equation can be written as follows:
\begin{equation}
\label{PsiuP}
\Psi(P) = \psi(P) u(P)\,,
\end{equation}
where the positive energy spinor satisfies
\begin{equation}
\bar u(P)(i\gamma\cdot P+M) = 0=(i\gamma\cdot P + M)u(P)\,,
\end{equation}
is normalised such that $\bar u(P) u(P) = 2 M$, and
\begin{equation}
2 M \Lambda_+(P) = \sum_{\sigma=\pm}u(P;\sigma)\bar u(P;\sigma) = M-i\gamma\cdot P\,,
\end{equation}
where in this line we have made the spin label explicit.  (See Ref.\,\cite[Appendix~A]{Chen:2012qr} for more details.)  Using Eq.\,\eqref{PsiuP}, then the complete SCI solution for $\psi(P)$ is a sum of the following Dirac structures ($\hat P^2=-1$):
\begin{equation}
\label{SAPD}
\psi^{\mathpzc S}(P) = {\mathpzc s}\,\mathbf{I}_{\rm D}\,,\;
\psi^{\mathpzc A}_{\mu}(P) = {\mathpzc a}_{1}\,i\gamma_{5}\gamma_{\mu}+{\mathpzc a}_{2}\gamma_{5}\hat P_{\mu}\,.
\end{equation}
As usual, $\bar \Psi(P) = \Psi(P)^\dagger \gamma_4 = \bar u(P) \gamma_4 \psi(P)^\dagger \gamma_4$.

Faddeev equation dynamics determines the values of the coefficients: $\{{\mathpzc s}, {\mathpzc a}_{1,2}\}$, each of which is a vector in flavour space.  The spin-flavour intertwining is determined by the quantum numbers of the baryon under consideration. Herein, we have the following structures:
{\allowdisplaybreaks
\begin{subequations}
\label{halfnucleon}
\begin{align}
\Psi_p & =
\left[
\begin{array}{ll}
{\rm r}_1 & u[ud]   \\
{\rm r}_2 &  d\{uu\}    \\
{\rm r}_3 &  u\{ud\}    \\
\end{array} \right],  \\
\Psi_n & =
\left[
\begin{array}{ll}
{\rm r}_1 & d[ud]   \\
{\rm r}_2 &  u\{dd\}    \\
{\rm r}_3 &  d\{ud\}    \\
\end{array} \right],  \\
\Psi_\Lambda & =
\frac{1}{\sqrt{2}}\left[
\begin{array}{ll}
{\rm r}_1 & -\sqrt{2}s[ud]           \\
{\rm r}_2 & u[ds]-d[us]     \\
{\rm r}_3 & u\{ds\} -d\{us\}
\end{array}
\right],  \label{SpinFlavourLambda}\\
\Psi_{\Sigma^+} & =
\left[
\begin{array}{ll}
{\rm r}_1 & u[us]   \\
{\rm r}_2 & s\{uu\} \\
{\rm r}_3 & u\{us\}\\
\end{array}
\right],\\
\Psi_{\Xi^0} & =
\left[
\begin{array}{ll}
{\rm r}_1 & s[us]   \\
{\rm r}_2 & s\{us\} \\
{\rm r}_3 & u\{ss\}\\
\end{array}
\right]. \label{BSXi0}
\end{align}
\end{subequations}
Since we work in the isospin symmetry limit, the $\Sigma^{0,-}$ and $\Xi^-$ structures may be obtained from those above by applying an isospin lowering operator.  These states are mass-degenerate with those written explicitly.}

\begin{table}[t]
\caption{\label{SolveFaddeev}
Masses and unit normalised Faddeev amplitudes obtained by solving the octet baryon Faddeev equations defined by Fig.\,\ref{figFaddeev}.
The row label superscript refers to Eqs.\,\eqref{halfnucleon}: for the $\Lambda$-baryon, $r_2$ is a scalar diquark combination; otherwise, it is axialvector.
Canonically normalised amplitudes, explained in connection with Eq.\,\eqref{CanonicalFA}, are obtained by dividing the amplitude entries in each row by the following numbers: ${\mathpzc n}_c^{p,n}=0.157$, ${\mathpzc n}_c^{\Lambda}=0.177$, ${\mathpzc n}_c^{\Sigma}=0.190$, ${\mathpzc n}_c^{\Xi}=0.201$.
(Masses listed in GeV.  Amplitudes are dimensionless. Recall that we work in the isospin-symmetry limit.)}
\begin{center}
\begin{tabular*}
{\hsize}
{
c@{\extracolsep{0ptplus1fil}}
|c@{\extracolsep{0ptplus1fil}}
c@{\extracolsep{0ptplus1fil}}
c@{\extracolsep{0ptplus1fil}}
c@{\extracolsep{0ptplus1fil}}
c@{\extracolsep{0ptplus1fil}}
c@{\extracolsep{0ptplus1fil}}
c@{\extracolsep{0ptplus1fil}}}\hline
  & mass
  & $s^{r_{1}}$ & $s^{r_{2}}$ & $a_{1}^{r_{2}}$ & $a_{2}^{r_{2}}$ & $a_{1}^{r_{3}}$ & $a_{2}^{r_{3}}$\\\hline
$p\ $ & $1.15$ & $\phantom{-}0.88$ & & $-0.38$ & $-0.063$ & $\phantom{-}0.27$ & $\phantom{-}0.044\ $  \\
$n\ $ & $1.15$ & $\phantom{-}0.88$ & & $\phantom{-}0.38$ & $\phantom{-}0.063$ & $-0.27$ & $-0.044\ $  \\
$\Lambda\ $ & $1.33$ & $\phantom{-}0.66$ & $0.62$ &  & & $-0.41$ & $-0.084\ $  \\
$\Sigma\ $ & $1.38$ & $\phantom{-}0.85$ & &$-0.46$  & $\phantom{-}0.15\phantom{3}$  & $\phantom{-}0.22$ & $\phantom{-}0.041\ $\\
$\Xi\ $ & $1.50$ & $\phantom{-}0.91$ & &$-0.29$  & $\phantom{-}0.021$  & $\phantom{-}0.29$ & $\phantom{-}0.052\ $
\\\hline
\end{tabular*}
\end{center}
\end{table}

Solving the Faddeev equations, one obtains the masses and amplitudes listed in Table~\ref{SolveFaddeev}.  The row labels therein refer to those identified in Eqs.\,\eqref{halfnucleon}.
Regarding the masses, we note that the values are deliberately $0.20(2)\,$GeV above experiment \cite{Zyla:2020zbs} because Fig.\,\ref{figFaddeev} describes the \emph{dressed-quark core} of each baryon.  To constitute a complete baryon, resonant contributions should be included in the Faddeev kernel.  Such ``meson cloud'' effects are known to lower the mass of octet baryons by $\approx 0.2$\,GeV \cite{Hecht:2002ej, Sanchis-Alepuz:2014wea}.  (Similar effects are reported in quark models \cite{Garcia-Tecocoatzi:2016rcj, Chen:2017mug}.)  Their impact on baryon structure can be estimated using dynamical coupled-channels models \cite{Aznauryan:2012ba, Burkert:2017djo}, but that is beyond the scope of contemporary Faddeev equation analyses.
Instead, we depict all form factors in terms of $x= Q^2/M_{B^\prime B}^2$, a procedure that has proved efficacious in developing sound comparisons with experiment \cite{Burkert:2017djo, Chen:2018nsg, Lu:2019bjs, Cui:2020rmu, ChenChen:2022qpy}.

Notwithstanding these remarks, the quark+diquark picture of baryon structure produces a $\Sigma-\Lambda$ mass splitting that is commensurate with experiment.  This is because the $\Lambda$ is primarily a scalar diquark system, whereas the $\Sigma$ has more axialvector strength: scalar diquarks are lighter than axialvector diquarks.

The Faddeev amplitudes in Table~\ref{SolveFaddeev} are unit normalised.  In calculating observables, one must use the canonically normalised amplitude.  That is defined via the baryon's Dirac form factor in elastic electromagnetic scattering, $F_1(Q^2=0)$.  To wit, for a baryon $B$, with $n_u$ $u$ valence-quarks, $n_d$ $d$ valence-quarks and $n_s$ $s$ valence-quarks, one decomposes the Dirac form factor as follows:
\begin{align}
& F_1^B(Q^2=0) \nonumber
\\ & = n_u e_u F_1^{Bu}(0) + n_d e_d F_1^{Bd}(0)+n_s e_s F_1^{Bs}(0)\,,
\label{CanonicalFA}
\end{align}
where $e_{u,d,s}$ are the quark electric charges, expressed in units of the positron charge.  It is subsequently straightforward to calculate the single constant factor that, when used to rescale the unit-normalised Faddeev amplitude for $B$, ensures $F_1^{Bu}(0)=1=F_1^{Bd}(0)=F_1^{Bs}(0)$.  So long as one employs a symmetry-preserving treatment of the elastic scattering problem, it is guaranteed that a single factor ensures all three flavour-separated electromagnetic form factors are unity at $Q^2=0$.  Explicit examples are provided elsewhere \cite{Wilson:2011aa}.

\subsection{Baryon currents}
\label{AppendixCurrents}
Using the propagators and amplitudes described above, one can write the explicit form of the baryon current indicated in Fig.\,\ref{figcurrent}.
Their content is most compactly expressed by associating a flavour-space column vector with the baryon spinor so that, \emph{e.g}., one may reexpress Eqs.\,\eqref{SAPD}, \eqref{BSXi0} as follows:
\begin{equation}
\underline{\Psi}_{\Xi^0} =
\Psi_{\Xi^0}^{{\mathpzc S}_{[us]}} {\mathpzc f}_s
+\Psi_{\Xi^0}^{{\mathpzc A}_{\{us\}}}{\mathpzc f}_s
+\Psi_{\Xi^0}^{{\mathpzc A}_{\{ss\}}}{\mathpzc f}_u\,,
\end{equation}
where ${\mathpzc f}_u={\rm column}[1,0,0]$, ${\mathpzc f}_d={\rm column}[0,1,0]$, ${\mathpzc f}_s={\rm column}[0,0,1]$.
The column vector that should be used is determined by $B$ and the specified diquark.  We denote the related row-vector by $\bar {\mathpzc f}_h$, $h=u,d,s$ and also define
\begin{equation}
\underline{S} = {\rm diagonal}[S_u, S_d, S_s]\,,
\end{equation}
where the quark propagators are drawn from Sec.\,\ref{AppendixSCI},

\subsubsection{Diagram 1}
This diagram expresses two contributions, Table~\ref{DiagramLegend}:
\begin{equation}
J_{5(\mu)}^1(K,Q) = J_{5(\mu)}^{qS}(K,Q)+J_{5(\mu)}^{qA}(K,Q)\,.
\end{equation}
Using the notation just introduced,
{\allowdisplaybreaks\begin{subequations}
\label{Diagram1Explicit}
\begin{align}
& J_{5(\mu)}^{qS}  =
\int_\ell \bar\Psi^{\mathpzc S}_{B^\prime}(P^\prime)\bar{\mathpzc f}_f \nonumber \\
& \quad \times  \underline S(\ell_{+}^\prime)
\underline\Gamma_{5(\mu)}^{fg}(Q)
\underline S(\ell_+)\Delta^{0^+}(-\ell) {\mathpzc f}_g \Psi^{\mathpzc S}_B(P), \\
& J_{5(\mu)}^{qA}  =
\int_\ell \bar\Psi^{\mathpzc A}_{B^\prime \alpha}(P^\prime) \bar{\mathpzc f}_f  \nonumber \\
& \quad \times \underline S(\ell_{+}^\prime)
\underline\Gamma_{5(\mu)}^{fg}(Q)
\underline S(\ell_+)\Delta_{\alpha\beta}^{1^+}(-\ell) {\mathpzc f}_g\Psi^{\mathpzc A}_{B\beta}(P),
\end{align}
\end{subequations}
where $\ell_\pm^{(\prime)}=\ell \pm P^{(\prime)}$,
the diquark propagators are given in Eqs.\,\eqref{qqPropagator},
and $\int_\ell$ represents our regularised four-dimensional momentum-space integral with, matching the Faddeev equation procedure, $\Lambda_{\rm uv}$ chosen to be the ultraviolet cutoff associated with the lightest diquark in the $B\stackrel{g\to f}{\to} B^\prime$ process.
}

The remaining elements in Eqs.\,\eqref{Diagram1Explicit} are
$\underline\Gamma_{5}^{fg} =: {\cal T}^{fg}\Gamma_{5}^{fg}$,
$\underline\Gamma_{5\mu}^{fg} =: {\cal T}^{fg}\Gamma_{5\mu}^{fg}$,
\emph{viz}.\ the dressed-quark+pseudoscalar, -quark+axialvector vertices that express the $g\to f$ quark transition.  Their calculation is exemplified in Ref.\,\cite[Eqs.\,(A.21)--(A.28)]{Xing:2022sor} and we adapt those results to all $g\to f$ transitions considered herein.
Notably, our implementation of the SCI guarantees the following (and other) Ward-Green-Takahashi identities ($k_+=k+Q$, $\underline{\mathpzc m}={\rm diagonal}[m_u,m_d,m_s]$):
\begin{align}
& Q_\mu \underline\Gamma_{5\mu}^{fg}(k_+,k)
+ i \underline{\mathpzc m}\,\underline\Gamma_{5}^{fg}(k_+,k)
+ i \underline\Gamma_{5}^{fg}(k_+,k) \underline {\mathpzc m} \nonumber \\
& = \underline S^{-1}(k_+)i\gamma_5 {\cal T}^{fg}
+ i\gamma_5 {\cal T}^{fg}\underline S^{-1}(k)\,. \label{WGTIfg}
\end{align}

\subsubsection{Diagram 2}
There is only one term in this case, \emph{i.e}., probe strikes axialvector diquark with dressed-quark spectator:
\begin{subequations}
\begin{align}
&J_{5(\mu)}^2(K,Q) = J_{5(\mu)}^{A^\prime A}(K,Q) \\
& = \int_\ell \bar\Psi^{{\mathpzc A}^\prime}_{B^\prime \alpha}(P^\prime) \bar{\mathpzc f}_h \underline S(\ell) \Delta_{\alpha\rho}^{1^+}(-\ell_-^\prime) \nonumber \\
& \times \Gamma^{A^\prime A}_{5(\mu),\rho\sigma}(-\ell_-^\prime,-\ell_-) \Delta_{\sigma\beta}^{1^+}(-\ell_-) {\mathpzc f}_h \Psi^{\mathpzc A}_{B\beta}(P),
\end{align}
\end{subequations}
where $\Gamma^{A^\prime A}_{5(\mu),\rho\sigma}$ is the axialvector diquark pseudoscalar (axialvector) vertex.  The associated form factors must be calculated; and to that end, we adapt the procedure detailed in Ref.\,\cite{Chen:2021guo}.  The results are collected in Appendix~\ref{AppendixCurrentsDiquark}, with those relevant here given in Eq.\,\eqref{GammaAA}.

\subsubsection{Diagram 3}
\label{AppendixDiagram3}
There are two terms in this case, \emph{i.e}., in the presence of a dressed-quark spectator, the probe strikes an axialvector (scalar) diquark, inducing a transition to a scalar (axialvector) diquark.  Writing the former explicitly:
\begin{subequations}
\begin{align}
&J_{5(\mu)}^3(K,Q) = J_{5(\mu)}^{SA}(K,Q) \\
& = \int_\ell \bar\Psi^{\mathpzc S}_{B^\prime}(P^\prime) \bar{\mathpzc f}_h\, \underline S(\ell) \Delta^{0^+}(-\ell_-^\prime) \nonumber \\
& \times \Gamma^{SA}_{5(\mu),\sigma}(-\ell_-^\prime,-\ell_-) \Delta_{\sigma\beta}^{1^+}(-\ell_-) {\mathpzc f}_h \Psi^{\mathpzc A}_{B\beta}(P),
\end{align}
\end{subequations}
where $\Gamma^{SA}_{5(\mu),\sigma}$ is the axialvector$\,\to\,$scalar diquark transition vertex.  Again, the associated form factors must be calculated, a task we complete following Ref.\,\cite{Chen:2021guo}.  The results are collected in Appendix~\ref{AppendixCurrentsDiquark}, with those relevant here given in Eq.\,\eqref{GammaSA}.  Naturally, $\Gamma^{AS}_{5(\mu),\sigma}(\ell^\prime,\ell)=-\Gamma^{SA}_{5(\mu),\sigma}(\ell^\prime,\ell)$.

\subsubsection{Diagram 4}
Here the probe strikes a dressed-quark ``in-flight'', emitted in the breakup of one diquark and en-route to formation of another:
\begin{align}
J_{5(\mu)}^4&(K,Q) = \sum_{J_1^{P_1}, J_2^{P_2}={\mathpzc S}, {\mathpzc A}}
\int_\ell \int_k
\bar\Psi^{J_2^{P_2}}_{B^\prime}(P^\prime) \bar{\mathpzc f}_{h^\prime} \Delta^{J_2^{P_2}}(k_{qq}) \nonumber \\
& \times \underline S(k) \underline\Gamma^{J_1^{P_1}}(\ell_{qq})
\left[\underline S(k_{qq}-\ell) \underline\Gamma_{5(\mu)}^{fg}(Q) \underline S(\ell_{qq}-k)\right]^{\rm T} \nonumber \\
& \times \bar{\underline\Gamma}^{J_2^{P_2}}(-k_{qq}) \underline S(\ell) \Delta^{J_1^{P_1}}(\ell_{qq})
{\mathpzc f}_h\Psi^{J_1^{P_1}}_{B}(P)\,, \label{DiagramFour}
\end{align}
where $(\cdot)^{\rm T}$ denotes matrix transpose,
$\bar \Gamma(K) = C^\dagger \Gamma(K)^{\rm T} C$,
and $\ell_{qq} = -\ell +P$, $k_{qq} = -k +P^\prime$.
We have suppressed Lorentz indices, which can readily be restored once the chosen transition is specified.

There are four terms in Eq.\,\eqref{DiagramFour}; but as exploited in the enumeration of Table~\ref{DiagramLegend}, symmetry relates ${\mathpzc S}{\mathpzc A}$ to ${\mathpzc A}{\mathpzc S}$; namely, there are only three distinct contributions.

It is worth highlighting here that in emulating the SCI formulation of the Faddeev equation in Ref.\,\cite{Chen:2012qr}, we have used a variant of the so-called ``static approximation'' \cite{Buck:1992wz}.  Consequently, the dressed-quark exchanged between the diquarks in the Faddeev kernel, Fig.\,\ref{figFaddeev}, is represented as
\begin{equation}
S^{\rm T}(q) \to \frac{g_B^2}{M_f}\,,
\label{staticexchange}
\end{equation}
with $g_B=1.18$.  Consistency with this simplification is achieved by writing
\begin{align}
& \underline S(k_{qq}-\ell) \underline\Gamma_{5(\mu)}^{fg}(Q) \underline S(\ell_{qq}-k) \nonumber \\
& \to \underline \Gamma_{5(\mu)}^{fg}(Q) g_B^2\left[\frac{1}{M_f}+\frac{1}{M_g}\right]
\frac{i\gamma\cdot Q + M_f + M_g}{Q^2+(M_f+M_g)^2}\,.
\end{align}

\subsubsection{Diagrams 5 and 6}
In a quark--plus--interacting-diquark picture of baryons, it is typically necessary to include ``seagull terms'' in order to ensure that relevant Ward-Green-Takahashi identities are satisfied \cite{Oettel:1999gc}.   Those relevant to the currents in Eqs.\,\eqref{jaxdq0}, \eqref{jpsdq0} are given in Ref.\,\cite{Chen:2021guo}.  Adapted to our SCI, they read
\begin{subequations}
\begin{align}
J_{5(\mu)}^5&(K,Q) = \sum_{J_1^{P_1}, J_2^{P_2}={\mathpzc S}, {\mathpzc A}}
\int_\ell \int_k
\bar\Psi^{J_2^{P_2}}_{B^\prime}(P^\prime) \bar{\mathpzc f}_{h^\prime} \Delta^{J_2^{P_2}}(k_{qq}) \nonumber \\
& \times \underline S(k) \chi_{5(\mu)}^{J_1^{P_1}fg}(\ell_{qq})
\underline S(k_{qq}-\ell)^{\rm T} \bar{\underline\Gamma}^{J_2^{P_2}}(-k_{qq})  \nonumber \\
& \times \underline S(\ell) \Delta^{J_1^{P_1}}(\ell_{qq})
{\mathpzc f}_h \Psi^{J_1^{P_1}}_{B}(P)\,, \label{DiagramFive}\\
J_{5(\mu)}^6&(K,Q) = \sum_{J_1^{P_1}, J_2^{P_2}={\mathpzc S}, {\mathpzc A}}
\int_\ell \int_k
\bar\Psi^{J_2^{P_2}}_{B^\prime}(P^\prime) \bar {\mathpzc f}_{h^\prime} \Delta^{J_2^{P_2}}(k_{qq}) \nonumber \\
& \times \underline S(k) \underline\Gamma^{J_1^{P_1}}(\ell_{qq})
\underline S(\ell_{qq}-k)^{\rm T}\bar\chi^{J_2^{P_2}fg}_{5(\mu)}(-k_{qq})  \nonumber \\
& \times \underline S(\ell) \Delta^{J_1^{P_1}}(\ell_{qq})
{\mathpzc f}_h\Psi^{J_1^{P_1}}_{B}(P)\,, \label{DiagramSix}
\end{align}
\end{subequations}
where, with $m_{P_{fg}}$ denoting the mass of the $f\bar g $ pseudoscalar meson,
{\allowdisplaybreaks
\begin{subequations}
\label{SeaGulls}
\begin{align}
\chi_{5\mu}^{J^{P} fg}(Q) & = - \frac{iQ_\mu}{Q^2+m_{P_{fg}}^2}
\left[\gamma_5 {\cal T}^{fg} \Gamma^{J^P}(Q)\right. \nonumber \\
& \quad \left. + \Gamma^{J^P}(Q) \left(\gamma_5 {\cal T}^{fg} \right)^{\rm T}\right] ,\\
i \chi_{5}^{J^{P} fg}(Q) & =
- \frac{1}{2{\mathpzc m}_{fg}}\frac{im_{P_{fg}}^2}{Q^2+m_{P_{fg}}^2}
\left[\gamma_5 {\cal T}^{fg} \Gamma^{J^P}(Q)\right. \nonumber \\
& \quad \left. + \Gamma^{J^P}(Q) \left(\gamma_5 {\cal T}^{fg} \right)^{\rm T}\right] ,\\
\bar\chi_{5\mu}^{J^{P} fg}(Q) & = - \frac{iQ_\mu}{Q^2+m_{P_{fg}}^2}
\left[\bar\Gamma^{J^P}(Q) \gamma_5 {\cal T}^{fg} \right. \nonumber \\
& \quad \left. + \left(\gamma_5 {\cal T}^{fg} \right)^{\rm T}\bar\Gamma^{J^P}(Q) \right] ,\\
i\bar\chi_{5}^{J^{P} fg}(Q) & = - \frac{1}{2{\mathpzc m}_{fg}}\frac{im_{P_{fg}}^2}{Q^2+m_{P_{fg}}^2}
\left[\bar\Gamma^{J^P}(Q) \gamma_5 {\cal T}^{fg} \right. \nonumber \\
& \quad \left. + \left(\gamma_5 {\cal T}^{fg} \right)^{\rm T}\bar \Gamma^{J^P}(Q) \right].
\end{align}
\end{subequations}
}

It is worth noting the following identity:
\begin{align}
Q_\mu& \chi_{5\mu}^{J^{P} fg}(Q)  + 2i{\mathpzc m}_{fg} \chi_{5}^{J^{P} fg}(Q) \nonumber \\
& = -i \gamma_5 {\cal T}^{fg} \Gamma^{J^P}(Q) - \Gamma^{J^P}(Q) \left(i\gamma_5 {\cal T}^{fg} \right)^{\rm T};  \label{WGTIfgseagull}
\end{align}
and the kindred relation for the conjugate seagulls.

\begin{figure}[!t]
\centerline{%
\includegraphics[clip, width=0.45\textwidth]{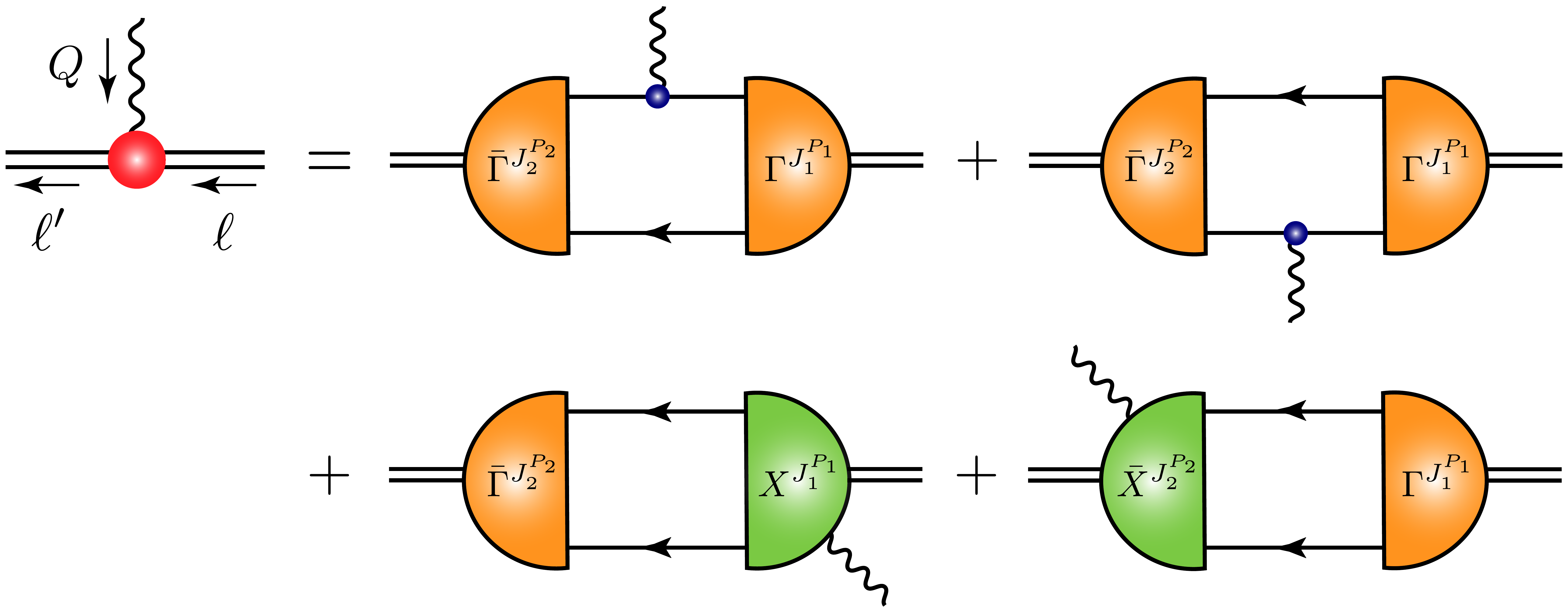}}
\caption{\label{figdqvx} Interaction vertex for the $J_1^{P_1}\to J_2^{P_2}$ diquark+probe interaction ($\ell^\prime=\ell+Q$): \emph{single line}, quark propagator; \emph{undulating  line}, pseudoscalar or axial current; $\Gamma$, diquark correlation amplitude; \emph{double line}, diquark propagator; and $\chi$, seagull interaction.}
\end{figure}

\subsection{Diquark currents}
\label{AppendixCurrentsDiquark}
In Appendix~\ref{AppendixCurrents} we saw that any study of baryon axial and pseudoscalar currents that exploits the quark+diquark representation of baryon structure requires knowledge of probe+diquark form factors.  We calculate them following the procedure detailed in Ref.\,\cite[Sec.\,III.C.4]{Chen:2021guo}, which employs the current depicted in Fig.\,\ref{figdqvx}.  Considering the systems involved, there are two form factors for each probe:  axialvector$\,\leftrightarrow\,$axialvector and axialvector$\,\leftrightarrow\,$pseudoscalar.

\subsubsection{Axialvector diquark transition form factors}
Using the SCI and considering $\{hg\}\to\{hf\}$ transition, the four diagrams in Fig.\,\ref{figdqvx} translate into the following expression:
{\allowdisplaybreaks
\begin{align}
\Gamma^{AA}_{5(\mu),\rho\sigma}&(\ell^\prime,\ell)  =
N_c^{\bar 3}{\rm tr}_{\rm DF}\int_t\left\{ i \bar{\underline\Gamma}_\rho^{\{h f\}}(-\ell^\prime)
\underline S(t_+^\prime)i\underline\Gamma_{5(\mu)}^{fg}(Q)\right. \nonumber \\
& \quad \times \underline S(t_+) i \underline\Gamma_\sigma^{\{g h\}}(\ell) \underline S(-t)^{\rm T} \nonumber \\
& + i \bar{\underline\Gamma}_\rho^{\{hf\}}(-\ell^\prime) \underline S(t) i \underline\Gamma_\sigma^{\{gh\}}(\ell) \nonumber \\
& \quad \times
\left[ \underline S(-t_-^\prime) i\underline \Gamma_{5(\mu)}^{fg}(Q) \underline S(-t_-) \right]^{\rm T} \nonumber \\
& -  i \bar{\underline \Gamma}_\rho^{\{h f\}}(-\ell^\prime) \underline S(t_+^\prime) \chi_{5(\mu),\sigma}^{\{gh\}fg}(\ell) \underline S(-t)^{\rm T} \nonumber \\
& \left. -  \bar\chi_{5(\mu),\rho}^{\{hf\}fg}(-\ell^\prime) \underline S(t_+) i\underline \Gamma_\sigma^{\{gh\}}(\ell) \underline S(-t)^{\rm T}\right\},
\label{GammaAA}
\end{align}
where we have made the Lorentz indices explicit, writing with reference to Eq.\,\eqref{DefineBSAs}, \emph{e.g}., $\underline\Gamma^{1^+}_{gh} = \underline\Gamma_\sigma^{\{g h\}}$;
$N_c^{\bar 3} = 2$ and the trace is over Dirac and flavour structure;
and $Q=\ell^\prime - \ell$,
$t_\pm^{(\prime)} = t \pm \ell^{(\prime)}$.
}

\subsubsection{Axialvector-scalar diquark transition form factors}
Analogously for the $\{hg\}\to[hf]$ transition, one has the following expression for the process described in Appendix~\ref{AppendixDiagram3}:
{\allowdisplaybreaks
\begin{align}
\Gamma^{SA}_{5(\mu),\sigma}&(\ell^\prime,\ell) =
N_c^{\bar 3}{\rm tr}_{\rm DF}\int_t \left\{
i \bar{\underline \Gamma}^{[hf]}(-\ell^\prime) \underline S(\ell_+^\prime)i\underline\Gamma_{5(\mu)}^{fg}(Q)
\right. \nonumber \\
& \quad \times \underline S(t_+) i \underline\Gamma_\sigma^{\{g h\}}(\ell) \underline S(-t)^{\rm T} \nonumber \\
& + i \bar{\underline\Gamma}^{[hf]}(-\ell^\prime) \underline S(t) i \underline\Gamma_\sigma^{\{gh\}}(\ell) \nonumber \\
& \quad \times
\left[ \underline S(-t_-^\prime) i\underline \Gamma_{5(\mu)}^{fg}(Q) \underline S(-t_-) \right]^{\rm T} \nonumber \\
& -  i \bar{\underline \Gamma}^{[h f]}(-\ell^\prime) \underline S(t_+^\prime) \chi_{5(\mu),\sigma}^{\{gh\}fg}(\ell) \underline S(-t)^{\rm T} \nonumber \\
& \left. -  \bar\chi_{5(\mu),\rho}^{[hf]fg}(-\ell^\prime) \underline S(t_+) i\underline \Gamma_\sigma^{\{gh\}}(\ell) \underline S(-t)^{\rm T}\right\}.
\label{GammaSA}
\end{align}
As noted above, $\Gamma^{AS}_{5(\mu),\sigma}(\ell^\prime,\ell)  = -\Gamma^{SA}_{5(\mu),\sigma}(\ell^\prime,\ell)$.
}

\begin{table}[t]
\caption{\label{interpolatorcoefficientsdu}
Probe-diquark form factors for $d\to u$ transitions, which for practical purposes can be interpolated using Eq.\,\eqref{interpolator12} with the coefficients listed here.  Where written, $f=d,u$ because we assume isospin symmetry; and the absence of an entry means the coefficient is zero.
(Every $\kappa(s)$ is dimensionless; so each coefficient in Eq.\,\eqref{interpolator12} has the mass dimension necessary to cancel that of the associated $s ({\rm GeV}^2)$ factor.)}
\begin{center}
\begin{tabular*}
{\hsize}
{
l@{\extracolsep{0ptplus1fil}}|
c@{\extracolsep{0ptplus1fil}}
c@{\extracolsep{0ptplus1fil}}
c@{\extracolsep{0ptplus1fil}}
c@{\extracolsep{0ptplus1fil}}}\hline
$\{fd\}\to\{fu\}\ $ & $a_0\ $ & $a_1\ $ & $b_1\ $ & $b_2\ $ \\
$\kappa_{\mathpzc p}^{AA}$ & $0.470\ $& $\phantom{-}0.173\ $ & $0.598\ $&   \\
$\kappa_{{\mathpzc a}_1}^{AA}$ & $0.467\ $& $\phantom{-}0.023\ $ & $0.598\ $&   \\
$\kappa_{{\mathpzc a}_2}^{AA}$ & $0.470\ $& $\phantom{-}0.023\ $ & $0.598\ $&   \\
$\kappa_{{\mathpzc a}_3}^{AA}$ & & & & \\\hline
$\{ds\}\to\{us\}\ $ & $a_0\ $ & $a_1\ $ & $b_1\ $ & $b_2\ $ \\
$\kappa_{\mathpzc p}^{AA}$ & $0.492\ $& $\phantom{-}0.137\ $ & $0.567\ $&   \\
$\kappa_{{\mathpzc a}_1}^{AA}$ & $0.489\ $& $-0.095\ $ & $0.444\ $& $-0.129\ $  \\
$\kappa_{{\mathpzc a}_2}^{AA}$ & $0.492\ $& $-0.096\ $ & $0.444\ $& $-0.129\ $ \\
$\kappa_{{\mathpzc a}_3}^{AA}$ & & & & \\\hline
$\{ff\}\leftrightarrow [ud]\ $ & $a_0\ $ & $a_1\ $ & $b_1\ $ & $b_2\ $ \\
$\kappa_{\mathpzc p}^{SA}$ & $0.649\ $& $\phantom{-}0.094\ $ & $0.182\ $&   \\
$\kappa_{{\mathpzc a}_1}^{SA}$ & $0.649\ $& $\phantom{-}0.327\ $ & $0.751\ $& $-0.035\ $  \\
$\kappa_{{\mathpzc a}_2}^{SA}$ & $0.646\ $& $\phantom{-}0.327\ $ & $0.751\ $& $-0.035\ $ \\\hline
$\{(u,d) s\}\leftrightarrow [(d,u) s]\ $ & $a_0\ $ & $a_1\ $ & $b_1\ $ & $b_2\ $ \\
$\kappa_{\mathpzc p}^{SA}$ & $0.641\ $& $\phantom{-}0.152\ $ & $0.327\ $&   \\
$\kappa_{{\mathpzc a}_1}^{SA}$ & $0.641\ $& $\phantom{-}0.254\ $ & $0.679\ $& $-0.031\ $  \\
$\kappa_{{\mathpzc a}_2}^{SA}$ & $0.638\ $& $\phantom{-}0.254\ $ & $0.679\ $& $-0.031\ $ \\\hline
\end{tabular*}
\end{center}
\end{table}

\subsubsection{Ward-Green-Takahashi identities}
It is worth remarking here that, using Eqs.\,\eqref{WGTIfg}, \eqref{WGTIfgseagull} and kindred relations, one may straightforwardly verify the following results:
{\allowdisplaybreaks
\begin{subequations}
\label{WGTIagain}
\begin{align}
0 & = Q_\mu \Gamma^{AA}_{5 \mu,\rho\sigma}(\ell^\prime,\ell) + i 2{\mathpzc m}_{fg} \Gamma^{AA}_{5,\rho\sigma}(\ell^\prime,\ell) \,,\\
0 & = Q_\mu \Gamma^{SA}_{5 \mu,\rho}(\ell^\prime,\ell) + i 2{\mathpzc m}_{fg}\Gamma^{SA}_{5,\rho}(\ell^\prime,\ell)\,.
\end{align}
\end{subequations}
These identities were established elsewhere \cite{Chen:2021guo}.  Being general, they can be used to constrain \emph{Ans\"atze} for the vertices involved.  Nevertheless, herein, we compute the SCI results directly.
}

\begin{table}[t]
\caption{\label{interpolatorcoefficientsus}
Probe-diquark form factors for $s\to u$ transitions, which can be interpolated using Eq.\,\eqref{interpolator12} with the coefficients listed here.  Where written, $f=d,u$ because we assume isospin symmetry; and the absence of an entry means the coefficient is zero.
(Every $\kappa(s)$ is dimensionless; so each coefficient in Eq.\,\eqref{interpolator12} has the mass dimension necessary to cancel that of the associated $s ({\rm GeV}^2)$ factor.)}
\begin{center}
\begin{tabular*}
{\hsize}
{
l@{\extracolsep{0ptplus1fil}}|
c@{\extracolsep{0ptplus1fil}}
c@{\extracolsep{0ptplus1fil}}
c@{\extracolsep{0ptplus1fil}}
c@{\extracolsep{0ptplus1fil}}}\hline
$\{fs\}\to\{fu\}\ $ & $a_0\ $ & $a_1\ $ & $b_1\ $ & $b_2\ $ \\
$\kappa_{\mathpzc p}^{AA}$ & $0.516\ $& $\phantom{-}0.131\ $ & $0.482\ $&   \\
$\kappa_{{\mathpzc a}_1}^{AA}$ & $0.480\ $& $-0.087\ $ & $0.318\ $& $-0.096\ $  \\
$\kappa_{{\mathpzc a}_2}^{AA}$ & $0.516\ $& $-0.093\ $ & $0.325\ $& $-0.095\ $  \\
$\kappa_{{\mathpzc a}_3}^{AA}$ & $0.128\ $& $-0.019\ $ & $0.416\ $& $-0.089\ $  \\\hline
$\{ss\}\to\{us\}\ $ & $a_0\ $ & $a_1\ $ & $b_1\ $ & $b_2\ $ \\
$\kappa_{\mathpzc p}^{AA}$ & $0.519\ $& $\phantom{-}0.113\ $ & $0.496\ $&   \\
$\kappa_{{\mathpzc a}_1}^{AA}$ & $0.481\ $& $\phantom{-}1.807\ $ & $4.328\ $& $\phantom{-}2.142\ $  \\
$\kappa_{{\mathpzc a}_2}^{AA}$ & $0.519\ $& $\phantom{-}1.877\ $ & $4.188\ $& $\phantom{-}2.083\ $  \\
$\kappa_{{\mathpzc a}_3}^{AA}$ & $0.076\ $& $\phantom{-}0.183\ $ & $3.090\ $& $\phantom{-}1.657\ $  \\\hline
$\{ds\}\rightarrow [ud]\ $ & $a_0\ $ & $a_1\ $ & $b_1\ $ & $b_2\ $ \\
$\kappa_{\mathpzc p}^{SA}$ & $0.742\ $& $\phantom{-}0.173\ $ & $0.304\ $&   \\
$\kappa_{{\mathpzc a}_1}^{SA}$ & $0.742\ $& $\phantom{-}0.248\ $ & $0.568\ $& $-0.023\ $  \\
$\kappa_{{\mathpzc a}_2}^{SA}$ & $0.712\ $& $\phantom{-}0.246\ $ & $0.552\ $& $-0.023\ $ \\\hline
$\{ss\}\rightarrow [us]\ $ & $a_0\ $ & $a_1\ $ & $b_1\ $ & $b_2\ $ \\
$\kappa_{\mathpzc p}^{SA}$ & $0.691\ $& $\phantom{-}0.179\ $ & $0.376\ $&   \\
$\kappa_{{\mathpzc a}_1}^{SA}$ & $0.691\ $& $\phantom{-}0.199\ $ & $0.547\ $& $-0.024\ $  \\
$\kappa_{{\mathpzc a}_2}^{SA}$ & $0.666\ $& $\phantom{-}0.195\ $ & $0.527\ $& $-0.023\ $ \\\hline
$\{fs\}\rightarrow [uf]\ $ & $a_0\ $ & $a_1\ $ & $b_1\ $ & $b_2\ $ \\
$\kappa_{\mathpzc p}^{SA}$ & $0.651\ $& $\phantom{-}0.144\ $ & $0.301\ $&   \\
$\kappa_{{\mathpzc a}_1}^{SA}$ & $0.651\ $& $\phantom{-}0.242\ $ & $0.574\ $& $-0.024\ $  \\
$\kappa_{{\mathpzc a}_2}^{SA}$ & $0.630\ $& $\phantom{-}0.238\ $ & $0.556\ $& $-0.023\ $ \\\hline
\end{tabular*}
\end{center}
\end{table}

\begin{table}[t]
\caption{\label{interpolatorcoefficientsneutral}
Probe-diquark form factors for $g\to g$, $g=u,d,s$, neutral current transitions, which can be interpolated using Eq.\,\eqref{interpolator12} with the coefficients listed here.  Where written, $f=d,u$ because we assume isospin symmetry; and the absence of an entry means the coefficient is zero.  \emph{N.B}.\ $\kappa_{{\mathpzc a}_3}^{AA}\equiv 0$ in this case.
(Every $\kappa(s)$ is dimensionless; so each coefficient in Eq.\,\eqref{interpolator12} has the mass dimension necessary to cancel that of the associated $s ({\rm GeV}^2)$ factor.)}
\begin{center}
\begin{tabular*}
{\hsize}
{
l@{\extracolsep{0ptplus1fil}}|
c@{\extracolsep{0ptplus1fil}}
c@{\extracolsep{0ptplus1fil}}
c@{\extracolsep{0ptplus1fil}}
c@{\extracolsep{0ptplus1fil}}}\hline
$\{ff\}\to\{ff\}\ $ & $a_0\ $ & $a_1\ $ & $b_1\ $ & $b_2\ $ \\
$\kappa_{\mathpzc p}^{AA}$ & $0.470\ $& $\phantom{-}0.173\ $ & $0.598\ $&   \\
$\kappa_{{\mathpzc a}_1}^{AA}$ & $0.467\ $& $\phantom{-}0.023\ $ & $0.598\ $& \\
$\kappa_{{\mathpzc a}_2}^{AA}$ & $0.470\ $& $\phantom{-}0.023\ $ & $0.598\ $& \\\hline
$\{ss\}\to\{ss\}\ $ & $a_0\ $ & $a_1\ $ & $b_1\ $ & $b_2\ $ \\
$\kappa_{\mathpzc p}^{AA}$ & $0.547\ $& $\phantom{-}0.094\ $ & $0.435\ $&   \\
$\kappa_{{\mathpzc a}_1}^{AA}$ & $0.475\ $& $\phantom{-}0.643\ $ & $1.878\ $& $\phantom{-}0.723\ $  \\
$\kappa_{{\mathpzc a}_2}^{AA}$ & $0.547\ $& $\phantom{-}0.654\ $ & $1.722\ $& $\phantom{-}0.649\ $   \\\hline
$\{fs\}\to\{fs\}\ $ & $a_0\ $ & $a_1\ $ & $b_1\ $ & $b_2\ $ \\
$\kappa_{{\mathpzc p}ff}^{AA}$ & $0.492\ $& $\phantom{-}0.137\ $ & $0.567\ $&   \\
$\kappa_{{\mathpzc a}_1ff}^{AA}$ & $0.489\ $& $-0.095\ $ & $0.444\ $& $-0.129\ $  \\
$\kappa_{{\mathpzc a}_2ff}^{AA}$ & $0.492\ $& $-0.096\ $ & $0.444\ $& $-0.129\ $   \\
$\kappa_{{\mathpzc p}ss}^{AA}$ & $0.564\ $& $\phantom{-}0.106\ $ & $0.416\ $&   \\
$\kappa_{{\mathpzc a}_1ss}^{AA}$ & $0.494\ $&  & $0.462\ $&  \\
$\kappa_{{\mathpzc a}_2ss}^{AA}$ & $0.564\ $&  & $0.469\ $&   \\\hline
$\{ud\}\leftrightarrow [ud]\ $ & $a_0\ $ & $a_1\ $ & $b_1\ $ & $b_2\ $ \\
$\kappa_{\mathpzc p}^{SA}$ & $0.649\ $& $\phantom{-}0.094\ $ & $0.182\ $&   \\
$\kappa_{{\mathpzc a}_1}^{SA}$ & $0.649\ $& $\phantom{-}0.327\ $ & $0.751\ $& $-0.035\ $  \\
$\kappa_{{\mathpzc a}_2}^{SA}$ & $0.646\ $& $\phantom{-}0.327\ $ & $0.751\ $& $-0.035\ $ \\\hline
%
%
%
$\{fs\}\leftrightarrow [fs]\ $ & $a_0\ $ & $a_1\ $ & $b_1\ $ & $b_2\ $ \\
$\kappa_{{\mathpzc p}ff}^{SA}$ & $0.641\ $& $\phantom{-}0.152\ $ & $0.327\ $&   \\
$\kappa_{{\mathpzc a}_1 ff}^{SA}$ & $0.641\ $& $\phantom{-}0.254\ $ & $0.679\ $& $-0.031\ $  \\
$\kappa_{{\mathpzc a}_2 ff}^{SA}$ & $0.638\ $& $\phantom{-}0.254\ $ & $0.679\ $& $-0.031\ $ \\
$\kappa_{{\mathpzc p}ss}^{SA}$ & $0.742\ $& $\phantom{-}0.160\ $ & $0.310\ $&   \\
$\kappa_{{\mathpzc a}_1 ss}^{SA}$ & $0.742\ $& $\phantom{-}0.186\ $ & $0.455\ $& $-0.018\ $  \\
$\kappa_{{\mathpzc a}_2 ss}^{SA}$ & $0.701\ $& $\phantom{-}0.185\ $ & $0.434\ $& $-0.017\ $ \\\hline
\end{tabular*}
\end{center}
\end{table}

\subsubsection{Probe-diquark form factors}
The expression in Eq.\,\eqref{GammaAA} yields the following explicit results:
{\allowdisplaybreaks
\begin{subequations}
\begin{align}
\Gamma^{AA}_{5,\rho\sigma}&(\ell^\prime,\ell)  =
- \frac{1}{2 {\mathpzc m}_{fg}} \frac{m_{P_{fg}}^2}{Q^2+m_{P_{fg}}^2}
\nonumber \\
&  \times
\varepsilon_{\alpha\beta\gamma\delta}\bar\ell_\gamma Q_\delta \kappa_{{\mathpzc p}fg}^{AA}(Q^2)T_{\rho\alpha}^{\ell^\prime}T_{\sigma\beta}^{\ell} \\
\Gamma^{AA}_{5\mu ,\rho\sigma}&(\ell^\prime,\ell)  =
\bigg[\varepsilon_{\alpha\beta\gamma\delta}\bar\ell_\gamma Q_\delta \frac{Q_\mu}{Q^2+m_{P_{fg}}^2}\kappa_{{\mathpzc a}_1fg}^{AA}(Q^2)\nonumber \\
&  + \varepsilon_{\mu\alpha\beta\gamma}[\bar\ell_\gamma \kappa_{{\mathpzc a}_2fg}^{AA}(Q^2)
+  Q_\gamma \kappa_{{\mathpzc a}_3fg}^{AA}(Q^2)]\bigg]T_{\rho\alpha}^{\ell^\prime}T_{\sigma\beta}^{\ell}\,,
\end{align}
\end{subequations}
where $ \bar\ell = \ell^\prime+\ell$ and, on the domain $Q^2 \in (-m_{P_{fg}}^2, 2M_{B^\prime B}^2)$ the computed form factors $\kappa_{{\mathpzc i}fg}^{AA}(Q^2)$, ${\mathpzc i}={\mathpzc p}, {\mathpzc a}_1,{\mathpzc a}_2,{\mathpzc a}_3$, are reliably interpolated using the following function:
\begin{equation}
\label{interpolator12}
\kappa (s=Q^2) = \frac{a_0 + a_1 s}{1+ b_1 s + b_2 s^2}\,,
\end{equation}
with the coefficients listed in Tables~\ref{interpolatorcoefficientsdu}, \ref{interpolatorcoefficientsus} (charged currents) and Table~\ref{interpolatorcoefficientsneutral} (neutral currents).  \emph{N.B}.\ Owing to the identities in Eqs.\,\eqref{WGTIagain}, $\kappa_{\mathpzc p}^{AA}(0)=\kappa_{{\mathpzc a}_2}^{AA}(0)$.  Moreover, in the isospin symmetry limit, $m_{\{fd\}}=m_{\{fu\}}$, $f=d,u$; consequently, $\kappa_{{\mathpzc a}_3ud}^{AA}\equiv 0$.  Furthermore, in no case considered herein does $\kappa_{{\mathpzc a}_3}^{AA}\neq 0$ contribute more than 1\% to any reported quantity.}

\begin{table}[thb]
\caption{\label{InterpolationsGA}
{\sf A}. Interpolation parameters for octet baryon axial transition form factors, Eq.\,\eqref{SCIinterpolationsGA}.
{\sf B}. Interpolation parameters for octet baryon induced pseudoscalar transition form factors, Eq.\,\eqref{SCIinterpolationsGP5}.
{\sf C}. Interpolation parameters for octet baryon pseudoscalar transition form factors, Eq.\,\eqref{SCIinterpolationsGP5R}.
(Every form factor is dimensionless; so each coefficient in Eq.\,\eqref{SCIinterpolationsGA} has the mass dimension necessary to cancel that of the associated $s ({\rm GeV}^2)$ factor.)}
\begin{center}
\begin{tabular*}
{\hsize}
{
l@{\extracolsep{0ptplus1fil}}
|c@{\extracolsep{0ptplus1fil}}
c@{\extracolsep{0ptplus1fil}}
c@{\extracolsep{0ptplus1fil}}
c@{\extracolsep{0ptplus1fil}}
c@{\extracolsep{0ptplus1fil}}}\hline
 {\sf A} & $g_0\ $ & $g_1\ $ & $g_2\ $ & $l_1\ $ & $l_2\ $ \\\hline
$\phantom{-}G_A^{pn}\ $ & $1.24\ $ & $1.97\ $ & $\phantom{-}0.29\phantom{0}\ $ & $2.44\ $ & $1.12\ $ \\
$\phantom{-}G_A^{\Lambda \Sigma^- }\ $ & $0.66\ $ & $1.19\ $ & $\phantom{-}0.16\phantom{0}\ $ & $2.73\ $ & $1.48\ $ \\
$-G_A^{p \Lambda}\ $ & $0.82\ $ & $1.00\ $ & $\phantom{-}0.074\ $ & $1.80\ $ & $0.68\ $ \\
$\phantom{-}G_A^{n\Sigma^- }\ $ & $0.34\ $ & $0.43\ $ & $\phantom{-}0.093\ $ & $1.86\ $ & $0.75\ $ \\
$\phantom{-}G_A^{\Sigma^+\Xi^0 }\ $ & $1.19\ $ & $3.28\ $ & $\phantom{-}0.33\phantom{0}\ $ & $3.35\ $ & $1.82\ $ \\
$\phantom{-}G_A^{\Lambda\Xi^- }\ $ & $0.23\ $ & $0.90\ $ & $-0.011\ $ & $4.42\ $ & $2.14\ $ \\
\hline
\end{tabular*}

\medskip

\begin{tabular*}
{\hsize}
{
l@{\extracolsep{0ptplus1fil}}
|c@{\extracolsep{0ptplus1fil}}
c@{\extracolsep{0ptplus1fil}}
c@{\extracolsep{0ptplus1fil}}
c@{\extracolsep{0ptplus1fil}}
c@{\extracolsep{0ptplus1fil}}}\hline
 {\sf B}  & $g_0\ $ & $g_1\ $ & $g_2\ $ & $l_1\ $ & $l_2\ $ \\\hline
$\phantom{-}G_P^{pn}\ $ & $2.01\ $ & $4.22\ $ & $\phantom{-}0.70\phantom{0}\ $ & $2.96\ $ & $1.57\ $ \\
$\phantom{-}G_P^{\Lambda \Sigma^- }\ $ & $1.25\ $ & $2.09\ $ & $\phantom{-}0.24\phantom{0}\ $ & $2.59\ $ & $1.25\ $ \\
$-G_P^{p \Lambda}\ $ & $1.18\ $ & $1.91\ $ & $\phantom{-}0.15\phantom{0}\ $ & $2.18\ $ & $0.80\ $ \\
$\phantom{-}G_P^{n\Sigma^- }\ $ & $0.50\ $ & $0.44\ $ & $\phantom{-}0.061\ $ & $1.39\ $ & $0.29\ $ \\
$\phantom{-}G_P^{\Sigma^+\Xi^0 }\ $ & $1.97\ $ & $2.38\ $ & $\phantom{-}0.060\ $ & $1.84\ $ & $0.43\ $ \\
$\phantom{-}G_P^{\Lambda\Xi^- }\ $ & $0.40\ $ & $1.34\ $ & $-0.014\ $ & $3.91\ $ & $1.88\ $ \\
\hline
\end{tabular*}

\medskip

\begin{tabular*}
{\hsize}
{
l@{\extracolsep{0ptplus1fil}}
|c@{\extracolsep{0ptplus1fil}}
c@{\extracolsep{0ptplus1fil}}
c@{\extracolsep{0ptplus1fil}}
c@{\extracolsep{0ptplus1fil}}
c@{\extracolsep{0ptplus1fil}}}\hline
{\sf C}  & $g_0\ $ & $g_1\ $ & $g_2\ $ & $l_1\ $ & $l_2\ $ \\\hline
$\phantom{-}G_5^{pn}\ $ & $1.24\ $ & $\phantom{-}0.13\phantom{0}\ $ & $\phantom{-}0.12\phantom{0}\ $ & $\phantom{-}0.19\ $ & $\phantom{-}0.13\phantom{0}\ $ \\
$\phantom{-}G_5^{\Lambda \Sigma^- }\ $ & $0.66\ $ & $\phantom{-}0.19\phantom{0}\ $ & $\phantom{-}0.075\ $ & $\phantom{-}0.36\ $ & $\phantom{-}0.18\phantom{0}\ $ \\
$-G_A^{p \Lambda}\ $ & $0.82\ $ & $\phantom{-}0.26\phantom{0}\ $ & $\phantom{-}0.14\phantom{0}\ $ & $\phantom{-}0.39\ $ & $\phantom{-}0.25\phantom{0}\ $ \\
$\phantom{-}G_5^{n\Sigma^- }\ $ & $0.34\ $ & $-0.13\phantom{0}\ $ & $\phantom{-}0.019\ $ & $-0.30\ $ & $\phantom{-}0.050\ $ \\
$\phantom{-}G_5^{\Sigma^+\Xi^0 }\ $ & $1.19\ $ & $\phantom{-}1.10\phantom{0}\ $ & $\phantom{-}0.26\phantom{0}\ $ & $\phantom{-}1.03\ $ & $\phantom{-}0.42\phantom{0}\ $ \\
$\phantom{-}G_5^{\Lambda\Xi^- }\ $ & $0.23\ $ & $\phantom{-}0.097\ $ & $-0.014\ $ & $\phantom{-}0.73\ $ & $-0.12\phantom{0}\ $ \\
\hline
\end{tabular*}
\end{center}
\end{table}

Turning to Eq.\,\eqref{GammaSA}, one finds:
{\allowdisplaybreaks
\begin{subequations}
\begin{align}
\Gamma^{SA}_{5,\rho}&(\ell^\prime,\ell)=
T_{\rho\alpha}^\ell Q_\alpha \frac{m_{P_{fg}}^2}{Q^2+m_{P_{fg}}^2} \nonumber \\
& \times \frac{m_{[hf]}+m_{\{gh\}}}{2{\mathpzc m}_{fg}} i \kappa_{{\mathpzc p}fg}^{SA}(Q^2) \,,\\
\Gamma^{SA}_{5\mu ,\rho}&(\ell^\prime,\ell)  =
T_{\rho\alpha}^\ell [m_{[hf]}+m_{\{gh\}}]
\bigg[ \delta_{\alpha\mu}
\kappa_{{\mathpzc a}_1fg}^{SA}(Q^2) \nonumber \\
&
- \frac{Q_\mu Q_\alpha}{Q^2+m_{P_{fg}}^2} \kappa_{{\mathpzc a}_2fg}^{SA}(Q^2)
\bigg]\,,
\end{align}
\end{subequations}
where the form factors can again be interpolated using Eq.\,\eqref{interpolator12} with the coefficients listed in Tables~\ref{interpolatorcoefficientsdu}\,--\,\ref{interpolatorcoefficientsneutral}.
}


\section{Interpolations of SCI Baryon Form Factors}
On $t=Q^2 \in (-m_{P_{fg}}^2, 2M_{B^\prime B}^2)$, SCI form factors can reliably be interpolated using the following functions:
{\allowdisplaybreaks
\begin{subequations}
\label{SCIinterpolations}
\begin{align}
G_A^{B^\prime B}(s) & = \frac{g_0+g_1 s+g_2 s^2}{1+l_1 s + l_2 s^2}\,,
\label{SCIinterpolationsGA}\\
G_{P,5}^{B^\prime B}(s) & = \frac{g_0+g_1 s+g_2 s^2}{1+l_1 s + l_2 s^2} {\cal R}(s)
\label{SCIinterpolationsGP5}\\
{\cal R}(s) & = \frac{m_{P_{fg}}^2}{s+m_{P_{fg}}^2}
\frac{M_{B^\prime B}}{{\mathpzc m}_{fg}}\,, \label{SCIinterpolationsGP5R}
\end{align}
\end{subequations}
with the coefficients listed in Tables~\ref{InterpolationsGA}A\,--\,C.
}


\end{document}